\documentclass[fleqn,usenatbib,useAMS]{mnras}

\usepackage{graphicx}	
\usepackage{amsmath}	
\usepackage{amssymb}
\usepackage{subfigure}

\newcommand{\pow}[1]{\ifmmode{}^{#1}\else ${}^{#1}$\fi}

\newcommand{\cm}{\,\ifmmode{{\mathrm{cm}}}\else cm\fi}

\newcommand{\ergps}{\,{\rm erg}\,{\rm s}\ifmmode{}^{-1}\else${}^{-1}$\fi}
\newcommand{\Mpch}{\,{\rm Mpc}\,\ifmmode h^{-1}\else $h^{-1}$\fi}
\newcommand{\snru}{\,\ifmmode{\mathrm{Myr}^{-1}}\else Myr${}^{-1}$\fi}
\newcommand{\kms}{\,\ifmmode{\mathrm{km}\,\mathrm{s}^{-1}}\else km\,s${}^{-1}$\fi\xspace}

\newcommand{\tcc}{\,\ifmmode{t_{\mathrm{cc}}}\else $t_{\mathrm{cc}}$\fi}
\newcommand{\lshatter}{\ifmmode{\ell_{\mathrm{shatter}}}\else $\ell_{\mathrm{shatter}}$\fi}
\def\gsim{\;\rlap{\lower 2.5pt
 \hbox{$\sim$}}\raise 1.5pt\hbox{$>$}\;}
\def\lsim{\;\rlap{\lower 2.5pt
   \hbox{$\sim$}}\raise 1.5pt\hbox{$<$}\;}

\usepackage[T1]{fontenc}
\usepackage{ae,aecompl}

\usepackage{newtxtext,newtxmath} 

\title[...]{L\'evy Flights and Leaky Boxes: Anomalous Diffusion of Cosmic Rays}

\author[Liang \& Oh]{
Naixin Liang, S. Peng Oh
\\
Department of Physics, University of California, Santa Barbara, CA 93106, USA}

\date{Draft from \today}

\begin{document}
\label{firstpage}
\maketitle

\begin{abstract}
In classical diffusion, particle step-sizes have a Gaussian distribution. However, in superdiffusion, they have power-law tails, with transport dominated by rare, long `L\'evy flights'. Similarly, if the time interval between scattering events has power-law tails, subdiffusion occurs. Both forms of anomalous diffusion are seen in cosmic ray (CR) particle tracking simulations in turbulent magnetic fields. They also likely occur if CRs are scattered by discrete intermittent structures, as recently suggested. Anomalous diffusion mimics a scale-dependent diffusion coefficient, with potentially wide-ranging consequences. However, the finite size of galaxies implies an upper bound on step-sizes before CRs escape. This truncation results in eventual convergence to Gaussian statistics by the central limit theorem. Using Monte-Carlo simulations, we show that this occurs in both standard finite-thickness halo models, or when CR diffusion transitions to advection or streaming-dominated regimes. While optically thick intermittent structures produce power-law trapping times and thus subdiffusion, `gaussianization' also eventually occurs on timescales longer than the maximum trapping time. Anomalous diffusion is a transient, and converges to standard diffusion on the (usually short) timescale of particle escape, either from confining structures (subdiffusion), or the system as a whole (superdiffusion). Thus, standard assumptions of classical diffusion are physically justified in most applications, despite growing simulation evidence for anomalous diffusion. However, if escape times are long, this is no longer true. For instance, anomalous diffusion in the CGM or ICM would change CR pressure profiles. Finally, we show the standard diagnostic for anomalous diffusion, $\langle d^2 \rangle \propto t^{\alpha}$ with $\alpha \neq 1$, is {\it not} justified for truncated L\'evy flights, and propose an alternative robust measure. 

\end{abstract}

\section{Introduction}

Diffuse gas in galaxies exchange energy and momentum with two relativistic fluids: photons and cosmic rays (CRs), both of which often have energy densities comparable to the thermal gas itself. In both cases, understanding the transport of relativistic particles is crucial to understanding their interaction with the thermal gas, as well as predicting observational signatures. Radiative transfer of light is very well-studied, and has solid underpinnings \citep{mihalas84}. By contrast, CR transport is much less understood (for recent reviews, see \citealt{zweibel17,ruszkowski23}). It is clear that CRs must scatter frequently in our Galaxy, since CR confinement times determined from spallation (the Boron-to-Carbon ratio; B/C) and radioactive decay ($^{10}\text{Be}/^{9}\text{Be}$ abundance) are orders of magnitude longer than the light travel time. However, the exact nature of the underlying scattering mechanism--usually believed to be small scale magnetic fluctuations, which cause CRs to slowly diffuse in pitch angle--is as yet unclear. Detailed comparison of the two currently dominant paradigms, gyroresonant scattering of CRs by Alfv\'en waves generated by the CRs (self-confinement), or scattering by compressive MHD modes which cascade from larger scales (extrinsic turbulence), do not fare well against Galactic observational constraints \citep{kempski21,hopkins21-CR-problems}. In light of these difficulties, alternative models where CRs impulsively scatter in pitch angle due to rare large angle bends in magnetic field lines \citep{lemoine23,kempski23} or against intermittent discrete structures \citep{butsky24-CRscatter}, have been proposed. CR transport unfortunately requires one to perform radiative transfer without a robust understanding of the underlying opacity. 

Given these uncertainties, the prevailing paradigm for CR transport bears re-examination. For instance, the standard fluid expression for the CR flux in the tightly coupled (i.e. strong scattering) limit is \citep{skilling71,breitschwerdt91}: 
\begin{equation}
\mathbf{F} = (\mathbf{v} + \mathbf{v}_{\rm s}) (E_c + P_c) - \kappa \nabla P_c, 
\label{eq:CRflux}
\end{equation}
where $\mathbf{v}$ is the gas velocity, $\mathbf{v}_{\rm s} = \mathbf{v}_{\rm A} \rm{sgn}(\mathbf{B} \cdot \nabla P_c)$ is the CR streaming velocity with Alfv\'en speed $\mathbf{v}_{\rm A}$, and $\kappa$ is the momentum averaged CR diffusion coefficient. This expression indicates that strongly scattered CRs advect with the gas ($\mathbf{v}[E_c+P_c]$), and stream with the Alfv\'en waves which scatter them\footnote{This term is only present in self-confinement models; it arises because particle-wave scattering is so strong that CRs are nearly isotropic in the Alfv\'en wave frame.} ($\mathbf{v}_{\rm s}[E_c+P_c]$). Finally, CRs also diffuse relative to the moving gas (and Alfv\'en waves, in self-confinement), since they are executing a random walk with finite mean free path when scattered by the magnetic fluctuations. In our Galaxy, while empirically based models are different as to the relative importance of CR advection and streaming, {\it all} models require rigidity-dependent diffusion to match observational constraints. 

In this paper, we examine the consequences of relinquishing the assumption of standard diffusive CR transport. Fick's law ($\mathbf{F} = - \kappa \nabla P_c$) arises if particle displacements are drawn from a Gaussian distribution\footnote{Strictly speaking, all that is required is that all moments of the displacement PDF are finite. From the Central Limit Theorem, a large sum of such random variables will converge to give a Gaussian distribution.}, as Einstein showed in his seminal work on Brownian motion. Standard diffusion has the property that the variance of particle displacements grows linearly with time, $\langle x^2 \rangle \propto t$. However, there are many systems where the observed variance $\langle x^2 \rangle \propto t^\gamma$, where $\gamma < 1$ (subdiffusion), or $\gamma > 1$ (superdiffusion), behavior dubbed fractional or anomalous diffusion. These include particle separation in turbulent diffusion \citep{richardson26}, transport in disordered or fractal media \citep{bouchaud90}, or even biological systems such as foraging behavior \citep{reynolds09}; see \citet{metzler00,dubkov08,henry10,uchaikin13} for reviews. \citet{uchaikin13} specifically considers the fractional diffusion of cosmic rays, but from a more abstract and mathematical standpoint. In general, subdiffusion corresponds to a time-fractional derivative\footnote{A fractional derivative $\partial^{m/n}$ corresponds to a derivative which, taken $n$ times, corresponds to the $m$th order derivative $\partial^{m}$.}, while superdiffusion corresponds to a space-fractional diffusion operator. In a particle basis, these correspond to distributions with power-law tails in time-increments or spatial step-size respectively. For example, the Cauchy distribution is a well-known example in astrophysics. The `fat tails' imply that the distribution has infinite variance, and this violates a key provision of the Central Limit theory. The sum of repeated draws from such a distribution does not converge to a Gaussian. Instead, from the Generalized Central Limit Theorem \citep{gnedenko54}, the sum converges to a stable distribution (or L\'evy alpha-stable distribution), which also has power-law tails. Standard diffusion is well characterized by size-steps of order the mean free path. By contrast, the power-law tails in superdiffusion mean that particle transport is dominated by rare large jumps. 

In this paper, we focus on superdiffusion of CRs, although we also consider subdiffusion (\S\ref{subsec:subdiffusion}). There are several physical scenarios which can lead to CR fractional diffusion: 
\begin{itemize}
\item{{\it Magnetic field line wandering.} In MHD turbulence, B-field lines separate as $\Delta l \propto d^{3/2}$, where the distance $d$ is measured along B-field lines \citep{lazarian99}, with close parallels to Richardson diffusion \citep{eyink11}. This rapid field line divergence induces super-diffusive CR transport perpendicular to field lines, even if they propagate via standard diffusion parallel to field lines \citep{yan08}, as seen in test particle simulations \citep{xu13}. In some regimes, perpendicular subdiffusion has also been invoked \citep{giacalone99,kota00}. 
Note that numerical confirmation of superdiffusion employs particle tracing in a static snapshot of turbulent magnetic fields. Interestingly, simulations of CR streaming along turbulent magnetic fields with {\it no} perpendicular diffusion (apart from small numerical diffusion) still finds super-diffusive transport, $\langle \Delta l^2 \rangle \propto t^{4/3}$ \citep{sampson22}.
Since MHD turbulence is ubiquitous, we regard it as the most robust means of producing CR superdiffusion.}

\item{{\it Scattering by intermittent structures.} The difficulties of standard theories which rely on weak magnetic fluctuations $\delta B/B \ll 1$ to scatter CRs, in reproducing the  rigidity dependence of CR confinement times and scattering rates, has prompted suggestions that CRs are scattered by rare intermittent structures, perhaps with $\Delta B/B \sim 1$, which produce strong large-angle scattering \citep{lemoine23,kempski23,butsky24-CRscatter}. In contrast to standard models, where the mean free path $\lambda \sim r_{\rm g}/(\delta B/B)^2$ produces gradual diffusion in pitch angle, in such models the mean free path $\lambda \sim 1/(n \sigma)$ is simply set by the abundance $n$ and cross-section $\sigma$ of scattering patches. The observed rigidity dependence can be reproduced if rarer, larger structures scatter a broader range of CR rigidities \citep{butsky24-CRscatter}. Similarly, very small scale structures could be responsible for radio wave scattering in the ISM \citep{stanimirovic18}, potentially constraining CR scattering \citep{kempski24}. If the spatial distribution of scattering structures has a fractal geometry, or if the scattering structures are associated with MHD turbulence, then the attendant self-similarity implies a power-law distribution of scattering lengths. As an example, if CRs are scattered by intermittent field line reversals, it has been found in simulations that the PDF of field line curvature has a power-law tail \citep{lemoine23,kempski23}. Also, if CRs become `trapped' in the intermittent structures, this can introduce stochastic time delays between particle flights, leading to subdiffusion. At this point, CR scattering by intermittent structures is fairly speculative. However, we note that L\'evy flights for radio wave propagation due to non-Gaussian, intermittent density fluctuations has already been proposed, and can potentially explain puzzling scalings of pulse broadening with distance, as well as observed angular broadening profiles \citep{boldyrev03,boldyrev05}.

\item{{\it Magnetic Mirroring.} Magnetic mirroring of CRs between compressions due to slow and fast modes, combined with the superdiffusive separation of field lines in Alfv\'enic turbulence, can give rise to a new type of diffusion dubbed ``mirror diffusion" \citep{lazarian21,zhang24}. It has been argued that this can give rise to a new form of CR superdiffusion.}}

\end{itemize}

What is at stake? Modifications to CR transport obviously affect the entire gamut of CR phenomenology. For instance, the different CR pressure gradients which develop could affect the efficacy and spatial distribution of gas acceleration and heating in galactic winds. In terms of transport time, superdiffusive transport is roughly equivalent to a standard diffusion coefficient which increases with lengthscale (see \S\ref{sec:physics}). This obviously impacts predicted observational signatures, such as the CR gradient problem\footnote{The CR gradient problem is a discrepancy between standard diffusion model predictions of gamma-ray emission in our Galaxy, which are peaked toward the galactic center due to centrally concentrated star formation and CR production, compared to observations, which find significantly flatter profiles as a function of radius.} and the CR anisotropy problem\footnote{The CR anisotropy problem comes from the fact that observations differ from standard model predictions of the energy dependence of the CR dipole anisotropy, as well as the small-scale power associated with higher order multipoles.}. See \citet{gabici19} for a recent review of such issues. At the same time, all physical systems are finite, and finite physical cutoffs mean that all higher order moments are finite, not infinite. After a sufficiently long time (i.e. many scatterings), the Central Limit Theorem will hold, the particle displacement distribution becomes Gaussian, and we revert to standard diffusion. The conditions for superdiffusion to revert to standard diffusion \citep{mantegna94} must be studied carefully in the context of CR propagation. This turns out to be the crux of our paper. 

The outline of this paper is as follows. In \S\ref{sec:physics}, we review relevant physics of fractional diffusion, and present our Monte-Carlo simulation method. In \S\ref{sec:results}, we present the results of fractional diffusion in a variety of set-ups, including their interpretation. In \S\ref{sec:discussion}, we discuss implications, and conclude in \S\ref{sec:conclusion}. 

\section{Physics of Fractional Diffusion}
\label{sec:physics}

\subsection{Fractional Diffusion and Stable Distributions}
\label{sec:stable}

Most of the time, our conception of diffusion and random walks begins and ends with Brownian motion. In 1D Brownian motion, the PDF $P(x,t)$ of a particle's position is governed by the diffusion equation: 
\begin{equation}
\frac{\partial}{\partial t} P(x,t)= \kappa \frac{\partial^2}{\partial \, x^2} P(x,t)
\label{eq:standard-diffusion}
\end{equation}
where the diffusion coefficient $\kappa \equiv {\lim}_{\Delta t \rightarrow 0} \langle \Delta x^2/2\Delta t \rangle$ is the rate of change of the variance of $x$. For diffusion from initial conditions where all particles reside at the origin ($P(x,0)=\delta(x)$, the Green's function solution is the Gaussian PDF: 
\begin{equation}
P(x,t) = \frac{1}{\sqrt{4\pi \kappa t}} {\rm exp} \left(- \frac{x^2}{4 \kappa t} \right) 
\label{eq:gaussian}
\end{equation}
whose variance increases linearly with time, $\langle x^2 \rangle = 2 \kappa t$. The celebrated Einstein-Stokes relation $\kappa = k_{\rm B} T/(6 \pi \eta a)$, where $\eta$ is the fluid viscosity and $a$ is the particle radius, relates the macroscopic diffusion coefficient to microscopic particle motions. Indeed, we can understand the macroscopic Green's function (equation \ref{eq:gaussian}) from microscopic particle motions. The total displacement in a random walk is simply a sum of smaller random steps, and the distribution of this sum is the convolution of the distributions of individual steps. Thus, the PDF $P(x,t)$ must remain `stable' (i.e., self-similar up to shifts in origin and scale) under repeated convolution. As is well known, a Gaussian convolved with another Gaussian still gives a Gaussian. In fact, the {\it only} distribution of finite variance which gives a scaled version of the original distribution under convolution with itself is a Gaussian. This is the crux of the Central Limit Theorem, and the Green's function (equation \ref{eq:gaussian}) for the diffusion equation can be viewed as a straightforward consequence: we are simply summing independent random variables drawn from distributions with finite variance. However, as we shall see, the Gaussian is only a special case of L\'evy-stable distributions, which are the attractor solution for a PDF which undergoes repeated convolution with itself. 

Standard diffusion applies in an extremely broad array of physical systems, for the same reasons as the widespread applicability of the Central Limit Theorem. However, it is not the whole story. The first sign was the empirical finding that particle separations in a turbulent medium obey $\langle x^2 \rangle \propto t^3$ \citep{richardson26}, much faster than standard diffusion $\langle x^2 \rangle \propto t$. Later, it was found that similar `anomalous diffusion' across a broad range of physical systems can be captured by the more general fractional diffusion equation \citep{klages08}: 
\begin{equation}
\frac{\partial^{\beta}}{\partial t^{\beta}} P(x,t) = \kappa_{\rm \alpha,\beta} \frac{\partial^{\alpha}}{\partial x^{\alpha}} P(x,t)
\label{eq:fractional-diffusion}
\end{equation}
where in general $\alpha, \beta$ are not integers but fractions. Note that $\kappa_{\alpha,\beta}$ is a generalized diffusion coefficient with dimensions $[\kappa_{\alpha,\beta}] = L^{\alpha}/t^{\beta}$. Consistent with dimensional analysis, for fractional diffusion, the variance scales as 
\begin{equation}
\langle x^2 \rangle \sim (\kappa_{\alpha,\beta} t)^{2\beta/\alpha}. 
\label{eq:xsq}
\end{equation}
The fractional derivatives are defined via their Laplace or Fourier transforms\footnote{Nonetheless, for concreteness it's useful to consider derivatives of the monomial $x^n$. For integer $m$, the derivative $(d^m/dx^m) x^n = n!/(n-m)! \ x^{n-m}$. For non-integer $q$, the fractional derivative $(d^q/dx^q) x^p = \Gamma(1+p)/\Gamma(1+p-q) \ x^{p-q}$. Note that for fractional derivatives, the derivative of a constant (p=0) $(d^q/dx^q) x^0 \propto x^{-q}$ does not necessarily vanish.}. The time derivative is the Caputo fractional derivative, defined in Laplace space as: 
\begin{equation}
\mathcal{L}\left( \frac{\partial^{\beta}}{\partial t^{\beta}} P(x,t) \right) = s^{\beta}\tilde{P}(x,s) - s^{\beta -1} P(x,0)
\label{eq:laplace-time}
\end{equation}
where $\mathcal{L}$ is the Laplace transform $\mathcal{L}(P(x,t);s) = \int_0^{\infty} e^{-st} P(x,t) dt$, $s$ is the conjugate variable to time, and $\tilde{P}(x,s)$ is the Laplace transform of $P(x,t)$. The space derivative is the symmetric Riesz fractional derivative, defined in Fourier space as: 
\begin{equation}
\mathcal{F} \left( \frac{\partial^{\alpha}}{\partial x^{\alpha}} P(x,t) \right) = - |k|^\alpha \hat{P}(k,t)
\label{eq:space-derivative}
\end{equation}
where $\mathcal{F}$ denotes the Fourier transform, $k$ is the Fourier wavenumber, and $\hat{P}(k,t)$ is the spatial Fourier transform of $P(x,t)$. Non-rigorously, equation \ref{eq:space-derivative} can be viewed as a fractional version of the usual WKB Fourier substitution, $\alpha$, $\partial^{\alpha}/\partial \, x^{\alpha} \rightarrow k^{\alpha}$. Note that for $\beta=1$, equation \ref{eq:fractional-diffusion} reverts to the standard diffusion equation for $\alpha=2$. However, for $\alpha=1$, the fractional Laplacian differs from the usual first order spatial derivative. The step-sizes are distributed according to a Cauchy distribution, which is not equivalent to the standard advection equation. In particular, the fractional derivative is non-local and involve long-range interactions.  

\begin{figure}
\centering
\includegraphics[width=0.95\linewidth]{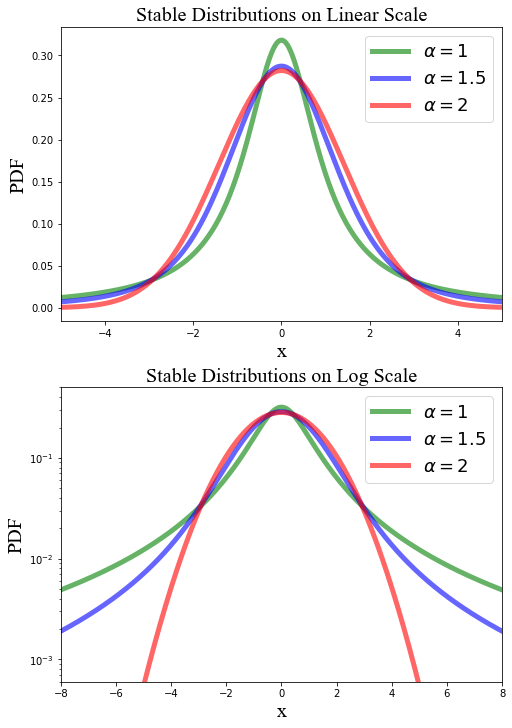}
\caption{\label{fig:pdfs}Stable distributions on (a) linear scale and (b) log scale, with $\alpha=1,1.5,2$ and same scale factor $\gamma=1$. While $\alpha=2$ gives the standard Gaussian distribution, $\alpha=1.5$ and $\alpha=1$ (the Cauchy distribution) have a stronger central peak (visible on the linear plot) and broad power law tails (visible on the log plot). In particular, they are significantly different at large $|x|$.}
\end{figure}

What is the Green's function for the fractional diffusion equation? It is useful to separate these into several distinct cases: 
\begin{itemize}
\item{{\it Standard diffusion} ($\beta=1, \alpha=2 $). Equivalent to equation (\ref{eq:standard-diffusion}), and therefore has a Gaussian Green's function (equation \ref{eq:gaussian}).}
\item{{\it Superdiffusion} ($\beta=1, 0<\alpha < 2 $). In this case, the time derivative is standard, and in particular the variance of the time interval between successive jumps is always finite. However, the distribution of step-sizes is drawn from a L\'evy-stable distribution, which is the most general PDF which is stable under the process of repeated convolution. These are given by the Fourier transform of an exponentiated power-law, which for symmetric distributions centered at the origin (the only form we consider) reads: 
\begin{equation}
P(x,t) = \mathcal{N} \int {\rm exp} (-ikx) {\rm exp} (-\gamma^{\alpha} |k|^{\alpha}) dk.
\label{eqn:L\'evy}
\end{equation}
where $\mathcal{N}$ is a normalization constant and the parameter $\gamma$ is a scale factor, which has the dimensions of length and sets the characteristic width of the distribution. This PDF is also the Green's function solution to the fractional diffusion equation (equation \ref{eq:fractional-diffusion}) when $\beta=1$, i.e. it gives the time-dependent distribution of particles $P(x,t)$ if they all start diffusing from the origin at $t=0$. In this case, the scale factor is: 
\begin{equation}
\gamma = (\kappa_{\alpha} t)^{1/\alpha}, 
\end{equation}
i.e. the characteristic width of the distribution $\langle \Delta x \rangle \propto t^{1/\alpha}$. In general, L\'evy-stable distributions are not analytic, with the well-known exceptions\footnote{And the less well-known exceptions of $\alpha=1/2$ (L\'evy-Smirnov) and $\alpha=3/2$ (Holtsmark).} of $\alpha=1$ (Cauchy) and $\alpha=2$ (Gaussian). The restriction to $0<\alpha\leq 2$ arises because $P_{\alpha}(x)$ is not positive definite for $\alpha > 2$, and cannot be normalized for $\alpha < 0$. 

The most crucial feature of L\'evy stable distributions is that they have an asymptotic power law tail: 
\begin{equation}
P_{\alpha} (x) \sim \frac{1}{x^{\alpha + 1}} \ \ \  \ \ \ (\alpha < 2). 
\label{eq:PL-tail}
\end{equation}
Intuitively, stable distributions must be self-similar (to retain the same shape under repeated convolutions), and power-laws are self-similar and indeed scale-free. This power-law tail also implies that all higher order moments $n \geq 2$, including the variance (n=2), diverge: 
\begin{equation}
\langle x^n \rangle = \int_0^{\infty} P_{\alpha} (x) x^n d\,x \sim x^{n-\alpha} \rightarrow \infty \ \ \  \ \ \ (n \geq 2; \alpha < 2).
\end{equation}
The variance of a stable distribution converges only when $\alpha=2$, the Gaussian case, which has exponential rather than power-law tails. Otherwise, the `fat' power-law tails of L\'evy-stable distributions mean that `rare' events are not vanishingly rare, and the term `L\'evy flight' denotes the fact that particle propagation is dominated by a few, very large steps, rather than the sum of many small steps, as in standard diffusion. In popular culture, the psychologically surprising, but mathematically expected occurrence of supposedly outlier events (which, for instance, play a key role in finance) has been popularized in `Black Swan Theory'.  

We show the stable distributions for $\alpha=1,1.5,2$ in Fig. \ref{fig:pdfs}, on both linear and log scales. While $\alpha=2$ gives the standard Gaussian distribution, $\alpha=1.5$ and $\alpha=1$ (the Cauchy distribution) have a stronger central peak (visible on the linear plot) and broad power law tails (visible on the log plot). The latter causes orders of magnitude differences at large $|x|$. Whilst this latter fact usually gets most of the attention, it is important to note that stable distributions with $\alpha < 2$ differs significantly with Gaussians {\it both} in the `core' (where they are more centrally peaked) and `halo' (where they fall off more slowly) parts of the distribution.
}

\item{{\it Subdiffusion} ($\alpha=2, 0< \beta < 1$). In standard random walks, we usually assume that the waiting time the scattering particle spends between each step is infinitesimally small. However, besides a distribution of step-sizes, there is also a distribution of scattering times, which in principle could be broad. This generally involves a situation where the scattering particle is trapped somewhere before taking the next step. Thus, a more generalized random walk can be generated by selecting scattering times from a probability distribution of times before each step, and then choosing the step lengths based on a probability distribution of step-sizes. For the standard time derivative, $\beta=1$, the distribution of times has a finite mean $\tau$. If so, for $t \gg \tau$ (so that $t^{-1} \sim s \ll \tau^{-1}$, or $\tau s \ll 1$), we can Taylor expand the Laplace transform $\tilde{P}(s) = 1- \tau s + \mathcal{O}(s^2)$) to obtain an exponential waiting time distribution: 
\begin{equation}
P(t) =  \frac{1}{\tau} {\rm exp \left( - \frac{t}{\tau} \right)}. 
\label{eq:time-exponential}
\end{equation}
Thus, standard diffusion (equation \ref{eq:standard-diffusion}) has an exponential waiting time distribution and a Gaussian distribution of pathlengths. In numerical simulations, we approximate the waiting time distribution with a fixed time interval $\tau$, so that $P(t) = \delta (t- \tau)$. 

However, instead of an exponential tail, there are also processes where the waiting time distribution has a power-law tail: 
\begin{equation}
P(t) \sim \frac{\beta \tau^{\beta}}{t^{1+\beta}} \ \ \ \ \ \  t > \tau, 0 < \beta < 1, 
\label{eq:time-PL}
\end{equation}
where $\tau$ is some timescale beyond which (for $t \gg \tau$) the distribution becomes power-law. Again, the distribution has `fat tails' compared to the standard case, but here in the waiting time distribution. Note that here, the mean waiting time $\int_0^{\infty} t P(t) \, dt$ diverges. This case is often called a `continuous time random walk', since there is no characteristic waiting timescale. The Green's function for subdiffusion is \citep{metzler00}: 
\begin{equation}
\label{eq:foxh}
\frac{1}{\sqrt{4\pi D t^\beta}} \, H^{2,0}_{1,2} \left( \frac{x^2}{4 D t^\beta} \Bigg| \begin{matrix} \left( 1 - \frac{\beta}{2}, \beta \right) \\ \left( 0, 1 \right), \left( \frac{1}{2}, 1 \right) \end{matrix} \right)
\end{equation}
where $H^{m,n}_{p,q}(z)$ is the Fox H function.
This special function can be evaluated numerically. 

The long tail in the waiting time distribution means that particles can get stuck in certain positions for very long times. The accumulation of particles at certain positions can create significant clumping which breaks ergodicity \citep{bel06}. The lack of scale separation between microscopic (single jump) and macroscopic ($P(x,t)$) timescales means that memory effects develop, and jumps are no longer Markov. Since $\langle x^2 \rangle \propto t^{\beta}$ where $0 < \beta < 1$, transport is slower than standard diffusion, and known as `subdiffusion'. Subdiffusion is observed in carrier transport in amorphous semiconductors \citep{pfister78}, granule transport in cells \citep{tolic04}, and chemical diffusion in aquifers \citep{scher02}. It is often associated with particle trapping. For instance, CRs could get trapped in magnetic mirrors, or other intermittent magnetic structures which scatter them. Intuitively, a power law distribution of structure sizes or magnetic curvature should lead to a power law distribution of waiting times. In \S\ref{subsec:subdiffusion}, we shall see that even intermittent scattering in structures of a fixed size can lead to a power law distribution of waiting times.}
\item{{\it General case.} The most general case allows both the step-size distribution $\lambda(x)$ and the waiting time distribution $\psi(t)$ to vary. If we continue to make the assumption that these two distributions are independent (see \S\ref{sec:convergence} for discussion of a case where this no longer holds), by Fourier Laplace transforms of the Chapman-Kolmogorov equation one obtains the propagator \citep{montroll65,klafter87}: 
\begin{equation}
P(k,s) = \frac{1 - \tilde{\psi}(s)}{s} \frac{1} {1-\hat{\lambda}(k)\tilde{\psi}(s)}, 
\label{eq:montroll-weiss}
\end{equation}
known as the Montroll-Weiss equation. This is equivalent to the fractional diffusion equation (equation \ref{eq:fractional-diffusion}), when $\tilde{\psi}(s)= {\rm exp}(-b s^{\beta}) \approx 1- b s^{\beta}$, and $\hat{\lambda}(k) = {\rm exp}(-a k^{\alpha}) \approx 1 - a k^{\alpha}$, as can be verified by using these Taylor expansions in equation \ref{eq:montroll-weiss}, and comparing against the Fourier Laplace transform of equation \ref{eq:fractional-diffusion}, and using equations \ref{eq:laplace-time}, \ref{eq:space-derivative}. Note that the exponentiated power-law form for $\tilde{\psi}(s),\hat{\lambda}(k)$ is required for them to be stable distributions with power-law tales (e.g., see equation \ref{eqn:L\'evy}). Since $\langle x^2 \rangle \sim t^{2\beta/\alpha}$, transport can either be super-diffusive ($\alpha < 2 \beta$), or sub-diffusive ($\alpha > 2 \beta$). The Green's function can also be expressed in terms of Fox H functions \citep{metzler00}.

As a practical matter, we do not solve the Montroll-Weiss or fractional diffusion equations directly in this paper, but rely on Monte-Carlo simulations (\S\ref{sec:monte}). We briefly mentioned them for completeness, since they are the fundamental governing equations.}
\end{itemize}

An important feature of solutions to the fractional diffusion equation is that they are self-similar. In the Green's function solutions to the diffusion equation, position $x$ and time $t$ do not appear separately, but always appear in the combination $x/(\kappa_{\alpha,\beta} t^{\beta})^{1/\alpha}$, so that it can be written in the form: 
\begin{equation}
G(x,t) = \frac{1}{(\kappa_{\alpha,\beta} t^{\beta})^{n/\alpha}} \psi (\frac{x}{(\kappa_{\alpha,\beta} t^{\beta})^{1/\alpha}}) 
\label{eq:G-self-similar}
\end{equation}
where $\psi$ is a stable distribution, and $n$ is the spatial dimension, i.e. the distribution rescales under the transformation $G(x,\Delta t) \rightarrow G(x/(\kappa_{\alpha,\beta} \Delta t^{\beta})^{n/\alpha},1)/(\kappa_{\alpha,\beta} \Delta t^{\beta})^{n/\alpha}$. For instance, comparing to equation \ref{eq:gaussian}, this clearly holds for standard diffusion ($\beta=1,\alpha=2$). Thus, just as in other self-similar situation (e.g., a Taylor-Sedov blast wave), for free diffusion, the solutions to the fractional diffusion equation can be rescaled to all lie on top of one another. This scaling behavior is useful for testing the accuracy of numerical solutions, even when we do not know the explicit form of the Green's function. It is also a useful way of confirming L\'evy flight dynamics. For instance, \citet{mantegna95} find that fluctuations in the S\&P 500 index from 1 min (then the minimum time to make a trade) to 1000 min intervals all have the same distribution, under suitable rescaling. 

\subsection{Convergence to Standard Diffusion}
\label{sec:convergence}

The `fat tails' of the waiting time distribution $\psi(t) \sim t^{-(1+\beta)}$ and step-size distribution $\lambda(x) \sim x^{-(1+\alpha)}$ cause the mean waiting time $\langle t \rangle \sim t^{1-\beta}$ and the variance in step-size $\langle x^2 \rangle \sim x^{2- \alpha}$ to diverge as $t\rightarrow \infty, x \rightarrow \infty$ for $0 < \beta < 1$, $0 < \alpha < 2$ respectively. The former produces subdiffusion, where particles can be frozen in place for long periods of time, while the latter produces superdiffusion, where particle transport is dominated by rare long flights. In practice, realistic physical systems always have temporal and spatial cutoffs: 
\begin{itemize}
\item{{\it Finite Jump Sizes.} Physical systems are finite in size, and so there must be a cutoff in jump size $l_{\rm max}$, beyond which the particle escapes the system. This implies that the power-law tail $P(x) \sim x^{-(1+\alpha)}$ of a L\'evy-stable distribution has a finite cutoff, with $P(x)=0$ for $|x| > l_{\rm max}$. If so, then all higher order moments are finite, and by the Central Limit Theorem, the sum of steps must converge to a Gaussian distribution, implying convergence to standard diffusion. The critical number of steps to converge to a Gaussian\footnote{Note that \citet{mantegna94} assumed $\gamma=1$ for the L\'evy-stable scale factor, and so derived $n_{\rm G} \propto l_{\rm max}^{\alpha}$. We generalize this to $n_{\rm G} \propto (l_{\rm max}/\gamma)^{\alpha}$, where $\gamma \sim (\kappa_{\alpha} \Delta t)^{1/\alpha}$.} is \citep{mantegna94}:
\begin{equation}
n_{\rm G} = A \left( \frac{l_{\rm max}}{(\kappa_{\alpha} \Delta t)^{1/\alpha}} \right)^{\alpha} = A \frac{l_{\rm max}^{\alpha}}{\kappa_{\alpha} \Delta t}
\label{eqn:ncrit} 
\end{equation}
where $\Delta t$ is the mean free time, and A is a constant:
\begin{equation}
A = {\left[\frac{\pi\alpha}{a\Gamma(1/\alpha){[\Gamma(1+\alpha)\sin(\pi\alpha/2)/(2-\alpha)]}^{1/2}}\right]}^{2\alpha/(\alpha-2)}. 
\label{eqn:const_A} 
\end{equation}
Thus, the critical number of steps $n_{\rm G}$ increases with both $l_{\rm max}$ and $\alpha$, as demonstrated in Fig. \ref{fig:pr}. This also implies a critical time for gaussianization:
\begin{equation}
t_{\rm G} \sim n_{\rm G} \Delta t \sim A \frac{l_{\rm max}^{\alpha}}{\kappa_{\alpha}},
\end{equation}
i.e., the time to `gaussianize' is of order the diffusion time across $l_{\rm max}$; once particles hit $t \gsim t_{\rm G}$, they have diffused a distance $x \gsim l_{\rm max}$, and the effects of the finite jump size become apparent. While \citet{mantegna94} obtained this result from a series expansion, this scaling can also be understood from the Berry-Esseen theorem \citep{shlesinger95}, which applies to symmetric random walks with finite second and third moments. It measures convergence to a normal distribution via the Kolmogorov-Smirnov distance, and states that for all $x$ and $n$ (where $n$ is the number of steps): 
\begin{equation}
|Q_{\rm n}(x) - N(x)| < \frac{5}{2} \frac{\langle |x|^3\rangle}{\langle x^2 \rangle^{3/2}} \frac{1}{\sqrt{n}}
\end{equation}
where $Q_{\rm n}(x)$ is the cumulative distance traveled after $n$ jumps, and $N(x)$ is the equivalent cumulative distance traveled if these jumps were drawn from a Gaussian distribution with the same variance. Since $\langle x^3 \rangle \propto l_{\rm max}^{3-\alpha}$, and $\langle x^2 \rangle^{3/2} \propto l_{\rm max}^{3(2-\alpha)/2}$, this gives $N \propto l_{\rm max}^{\alpha}$. 

Again, from dimensional analysis, once $n > n_{\rm G}$, and the system converges to standard diffusion, the effective diffusion coefficient is: 
\begin{equation}
\kappa_{\rm eff} \approx \kappa_{\alpha} l_{\rm max}^{2-\alpha}, 
\end{equation}
an ansatz we will test in our Monte-Carlo simulations (\S\ref{subsec:convergence}). Thus, early on, when 
$n < n_{\rm G}$, if they are erroneously assumed to undergo standard diffusion, this will appear as an apparent scale-dependent diffusion coefficient:
\begin{equation}
\kappa  \approx \kappa_{\rm eff} \left( \frac{l}{l_{\rm max}} \right)^{2-\alpha} = \kappa_{\alpha} l^{2-\alpha}
\label{eq:kappa_l}
\end{equation}

These ideas turn out to be pivotal for this paper, so we use Monte-Carlo simulations (\S\ref{sec:monte}) to demonstrate the convergence to Gaussian behavior. Naively, from equation \ref{eq:xsq} (and for $\beta=1$) for 1D diffusion we should see the variance in particle position $\langle x^{2} \rangle \propto t^{2/\alpha}$ transition to the Gaussian case $\langle x^{2} \rangle \propto t$ after time $t_{\rm G}$. However, as we later discuss in \S\ref{subsec:convergence}, this does not hold: instead, for truncated L\'evy flights, $\langle x^{2} \rangle \propto t$ even early on. Instead, we follow the analysis in \citet{mantegna94}, and use the fact that in 1D, the normalization of the L\'evy stable PDF scales as $t^{-1/\alpha}$ (equation \ref{eq:G-self-similar}), which is equivalent to the `probability of return', i.e. the probability $P_{\rm n}(0)$ that a particle returns to the origin $x=0$ after time $t$, or $n=t/\Delta t$ scatterings. Thus, we expect $P_{\rm n}(0) \propto n^{-1/\alpha}$ at early times, and $P_{\rm n}(0) \propto n^{-1/2}$ (the Gaussian scaling) after $n_{\rm G}$ scatterings.  To illustrate this, we initialize a delta source at the origin $z=0$ at $t=0$, allow particles to diffuse in 1D with $\alpha=1.5$, and estimate the probability of return. As shown in Fig. \ref{fig:pr}, $P_{\rm n}(0)$ indeed shows the expected change in scaling behavior after $n_{\rm G}$ steps. We find that the profile converges to standard diffusion with effective diffusion coefficient $\kappa_{\text{eff}}= C(\alpha) \kappa_{\alpha} l^{2-\alpha}$, where C is a constant.
\begin{figure}
\centering
\includegraphics[width=0.95\linewidth]{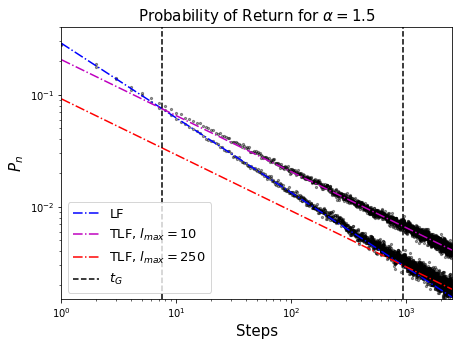}
\caption{\label{fig:pr}Monte-Carlo simulations of the probability of return $P_{n}(0)$, defined as the probability of $x_n=0$ after n steps. It is estimated by the fraction of particles close to x=0, within a scale factor $\gamma$. Both simulations have $\alpha=1.5$ but different path length truncations, $l=10$ and $l=250$. The small-n regime follows a L\'evy stable probability $P_{\rm n}(0) \propto n^{-1/\alpha}$, shown as a blue dashed line, and transitions to the Gaussian scaling at large $n$, $P_{\rm n}(0) \propto n^{-1/2}$. The crossover occurs at the number of steps predicted in Eq \ref{eqn:ncrit}, shown by vertical lines. Note the increase in Poisson noise for larger $n$.}
\end{figure}
}
\item{{\it Finite waiting times.} Particle trapping times are usually bounded -- there is almost always a finite probability for a particle to escape a trap, so that both the maximum trapping time $\tau_{\rm max}$, and the mean trapping time $\langle \tau \rangle \propto \tau_{\rm max}^{1-\beta}$ is always finite. For $t \ll \tau_{\rm max}$, the effect of the maximum trapping time is not apparent, and the system shows apparent subdiffusion. For $t \gg \tau_{\rm max}$, the mean waiting time is finite, and the system becomes ergodic. Trapping has little effect in the limit $t/\tau_{\rm max} \rightarrow \infty$, and transport converges to standard diffusion \citep{bouchaud90}. This transition between subdiffusion ($\langle x^2 \rangle \propto t^{\gamma}, \gamma < 1$) at early times and standard diffusion at late times ($\langle x^2 \rangle \propto t$) has been seen in experimental data \citep{saxton07}. The crossover between the two regimes is of order the maximum trapping time\footnote{Note therefore that the crossover time $t_{\rm c} \sim \tau_{\rm max}$ does {\it not} scale with the mean trapping time $\langle \tau \rangle$, since $\langle \tau \rangle \propto \tau_{\rm max}^{1-\beta}$; instead, $\tau_{\rm max} \propto \langle \tau \rangle^{1/(1-\beta)}$. 
} $\tau_{\rm max}$. From dimensional analysis, the effective diffusion coefficient once the system converges to standard ($\alpha=2$) diffusion is:
\begin{equation}
\kappa_{\rm eff} \sim \kappa_{\beta} \tau_{\rm max}^{\beta -1}, 
\end{equation}
an ansatz we will check in our Monte-Carlo simulations (\S\ref{subsec:subdiffusion}). If the system is incorrectly modeled using standard diffusion at all times, the initial transient subdiffusion will appear as a time-dependent diffusion coefficient: 
\begin{equation}
\kappa(t) \approx \kappa_{\rm eff} \left( \frac{\tau_{\rm max}}{t} \right)^{1-\beta}  = \frac{\kappa_{\beta}}{t^{1-\beta}} ; \ \ \ \ \ \ t < t_{\rm c}
\end{equation}
If convergence to standard diffusion is slow, it can be tricky to distinguish transient and true subdiffusion, though techniques exist for doing this \citep{berezhkovskii14}.} 

\item{While we have considered step function cutoffs $\tau_{\rm max},l_{\rm max}$ to the waiting time and jump-size distributions, equivalent results can be obtained for smooth (e.g., exponential) cutoffs, producing what are known as tempered stable distributions which are better-behaved analytically \citep{koponen95}.}
\end{itemize}

Since we expect all physical systems to have finite temporal and spatial cutoffs, it may seem surprising that we see fractional diffusion at all. Shouldn't all phenomenon simply converge to standard diffusion? The point is that convergence toward standard diffusion can be extremely slow, and generally requires large temporal or scale separations (so that either $t/\tau_{\rm max}$, or the number of scatterings $n$ is large) which may not hold. However, CR transport is a setting with a very wide range of scales, where it may indeed be realistic to expect such scale separation. We investigate this in detail in \S\ref{subsec:convergence}.

\subsection{L\'evy Walks: Finite Speed of Light Effects}

Thus far, we have ignored the fact that CRs have a maximum velocity, given by the speed of light. Even standard diffusion allows for infinite propagation velocities, which is clear from the Green's function solution (equation \ref{eq:gaussian}): even after an infinitesimally short time, there is a non-zero density of particles at arbitrarily large distances. However, the density of such fast particles (with $x > ct \gg (\kappa \Delta t)^{1/2}$) is exponentially suppressed. The `fat' power-law tails of L\'evy-stable distributions (equation \ref{eq:PL-tail}) means that the density of such super-luminal particles could be higher, and a potential source of concern. More generally, our treatment of fractional diffusion has assumed that the waiting time distribution $\psi(t)$ and the step-size distribution $\lambda(x)$ are independent. This need not be the case: for instance, longer waiting times could be correlated with subsequently larger jumps. Incorporating the finite speed of light is an example of the latter, since all particles must be confined within the ballistic cone $x=vt$, which couples intervals in space and time. Since the PDF is now bounded in space, all moments of the PDF are now finite, with potential implications for convergence to standard diffusion. Fractional diffusion where the finite speed of the random walker is taken into account is often dubbed `L\'evy walks' (see \citealt{zaburdaev15} for a comprehensive review), as opposed to `L\'evy flights'. 

In this paper, we ignore finite velocity effects, which are negligible for the scenarios we consider. Intuitively, since there is a large scale separation between the speed of light and typical CR diffusive ($v_{\rm D} \sim \kappa_{\alpha}/L^{\alpha-1}$), advective or streaming velocities, the finite speed of light has little effect\footnote{This fact is exploited in two-moment CR transport codes \citep{jiang18,thomas19,chan19} which use a `reduced speed of light' to increase the Courant time-step, and reduce computational cost.}. In a little more detail, we can justify it as follows. The propagator for super-diffusive L\'evy walks (which are more likely to be affected by the finite speed of light) is given by a L\'evy-stable distribution, sandwiched between delta function peaks at $x=-ct, x=ct$, corresponding to particles traveling along the ballistic cone. The characteristic width or scale factor $\gamma \sim (\kappa_{\alpha,\beta} t^{\beta})^{1/\alpha}$ has a sub-linear scaling with time for $\beta < \alpha$, which is generally true for the parameters we consider $0 < \beta \leq 1$, $1 < \alpha \leq 2$.
Thus, as time progresses, the influence of the ballistic cone wanes, since it expands faster than the scale factor ($x=ct$ vs. $\gamma \propto t^{\beta/\alpha}$, where $\beta/\alpha < 1$). 
However, if $\beta/\alpha >1 $, so that the scale factor expands faster with time than the ballistic cone, then the PDF becomes increasingly U-shaped, with most particles confined to the ballistic cone. Models with different time-space couplings are compared in \citet{uchaikin12}, which shows that the ballistic cone dominated case $\beta/\alpha > 1$ predicts unphysical results, such as excessive CR anisotropy. More to the point, long inferred confinement times imply that CRs do not travel ballistically. 

Although finite speed of light effects are unlikely to be important, other ways in which the waiting time and step-size distributions are coupled (e.g., via the topology of intermittent scattering structures) could be. However, this is a subject for future work, beyond the scope of this paper.

\subsection{Method: Monte-Carlo Simulations}
\label{sec:monte}

What is the best way to solve the fractional diffusion equation (equation \ref{eq:fractional-diffusion}) numerically? In principle it is possible to solve it on a grid code (e.g., see \citet{bonito18,lischke20} for reviews). However, this is still an area of active research, and there is currently no consensus on the best way to deal with what amounts to non-local transport, where grid cells far away provide non-negligible particle flux. There are two main approaches, which utilize either the spectral definition of fractional derivatives (which we have presented here, equations \ref{eq:laplace-time} and \ref{eq:space-derivative}), or the integral definition, which deals with singular non-integrable kernals. The latter can be solved via a matrix approach \citep{podlubny09}. The problem is non-trivial, not least because non-local boundary conditions (where the value of a function must be specified on the entire exterior of a domain) are required to compute the fractional Laplacian. 

For this reason, we use a simple Monte-Carlo approach. This is very popular in Ly$\alpha$ radiative transfer \citep{dijkstra19}, which faces the same problem of non-local photon transport due to long photon trajectories which arise when photons drift away from line center in frequency. Although it is noisy and low-order in accuracy\footnote{It is essentially equivalent to a first order Euler-Marayuma method for stochastic differential equations.}, and requires large numbers of particles and small time-steps to converge, it is straightforward to implement and robust.  Moreover, a particle based approach has several distinct advantages. For instance, it allows us to track the total age of CRs, and the time spent in various domains (disk, halo). It also straightforwardly captures the approach to `standard' diffusion if the flight path or waiting time of CRs is capped to finite values (\S\ref{sec:convergence}). This cannot easily be accomplished in a discretized code without additional machinery. 

In our Monte Carlo simulations, we track the motion of a large number of particles, representing CRs. In many cases, we continuously produce new particles at every time step, representing CR injection. CRs can also leave the system (`escape'), under conditions we will specify. Each particle is labeled with both an age $t$ (where $t=0$ when the particle is injected) and position $x$. We update these as follows: 
\begin{itemize}
\item{For standard time derivatives ($\beta=1$ in equation \ref{eq:fractional-diffusion}), we use a uniform time-step $\Delta t$ for all particles\footnote{In principle we should draw the time-step from an exponential distribution (equation \ref{eq:time-exponential}), but in practice this is not necessary; the difference between a step-function and exponential fall-off does not matter. Indeed, our Monte-Carlos converge to the correct analytic solutions for the $\beta=1$ case.}. We then draw the spatial jump $\Delta x$ from a symmetric L\'evy-stable distribution with stability parameter $\alpha$, using the SciPy stats package, with scale factor $\gamma=(\kappa_{\alpha} t)^{1/\alpha}$. We then update the ages $t \rightarrow t + \Delta t$ and positions $x \rightarrow x + \Delta x$ of all particles.}

\item{
For continuous time random walks ($0 < \beta < 1$), where particles undergo trapping, we draw a trapping time $\Delta t$ from a one-sided distribution with a power law tail (equation \ref{eq:time-PL}). In particular, the trapping time is chosen from the shifted Pareto law \citep{henry10}:
\begin{equation}
    \psi(t) = \frac{\beta/\tau}{(1 + t/\tau)^{1+\beta}}.
    \label{eq:pareto_law}
\end{equation}
The parameters $\beta$ and $\tau$ are the anomalous exponent and the characteristic timescale respectively. This probability density function has the asymptotic scaling
\begin{equation}
    \psi(t) \sim \frac{\beta}{\tau} \left( \frac{t}{\tau} \right)^{-1-\beta}
\end{equation}
for long times. A random waiting-time that satisfies the waiting-time density, Eq.~\eqref{eq:pareto_law}, can be generated as follows:
\begin{equation}
    \Delta t = \tau \left( \left( 1 - r \right)^{-\frac{1}{\beta}} - 1 \right)
\end{equation}
where $r \in (0, 1)$ is a uniform random number. We then repeat these steps until the total waiting time for all particles reach or exceed the required simulation time. Whenever the particle waiting time is updated, we also update the particle position $x \rightarrow x + \Delta x$, by drawing the jump length $\Delta x$ from a Gaussian distribution that correspond to standard spatial diffusion. This simulates the distance traveled by the particle after leaving the trapping structure and before entering another. Now the number of scattering steps is not proportional to the run time, but becomes a random variable that depends on the confining times in each trapping structure. To justify the usage of power-law distribution for waiting times, we simulate 1D and 3D spherical ‘patches’ where the particles are trapped. The time distribution which particles spent scattering inside the structure are characterized in \S\ref{subsec:subdiffusion}, and their scaling with the patch size is examined.} 

\item{In 3D simulations, we evolve each spatial dimension separately. Anisotropic diffusion can be taken into account by changing the diffusion coefficients in different directions.}

\item{We check convergence both with respect to the number of particles, and the simulation timestep $\Delta t$.}

\end{itemize}

\section{Simulation Results}
\label{sec:results}

We first consider superdiffusion in \S\ref{subsec:free}-\ref{subsec:energy}, where we vary $\alpha$ in equation \ref{eq:fractional-diffusion}, and draw step-sizes from a L\'evy-stable distribution with power-law tails. Otherwise, we set $\beta=1$, corresponding to the standard time derivative, and uniform time-increments $\Delta t$ in our Monte-Carlo simulations. We turn to subdiffusion -- where we vary $\beta$, and thus draw $\Delta t$ from a power-law distribution -- in \S\ref{subsec:subdiffusion}. To begin, we contrast `free diffusion', where CRs are allowed to diffuse in a quasi-infinite medium (\S\ref{subsec:free}), with diffusion in a bounded medium (\S\ref{subsec:absorb}), where absorbing barriers are placed at some finite distance from a CR source. This mimics the process of cosmic ray escape from the Galaxy. Subsequent sections all incorporate some form of absorption or CR escape.  

\subsection{Free Diffusion}
\label{subsec:free}

\subsubsection{Free Diffusion in 1D}
\label{subsec:1D}

\begin{figure}
\centering
\includegraphics[width=0.95\linewidth]{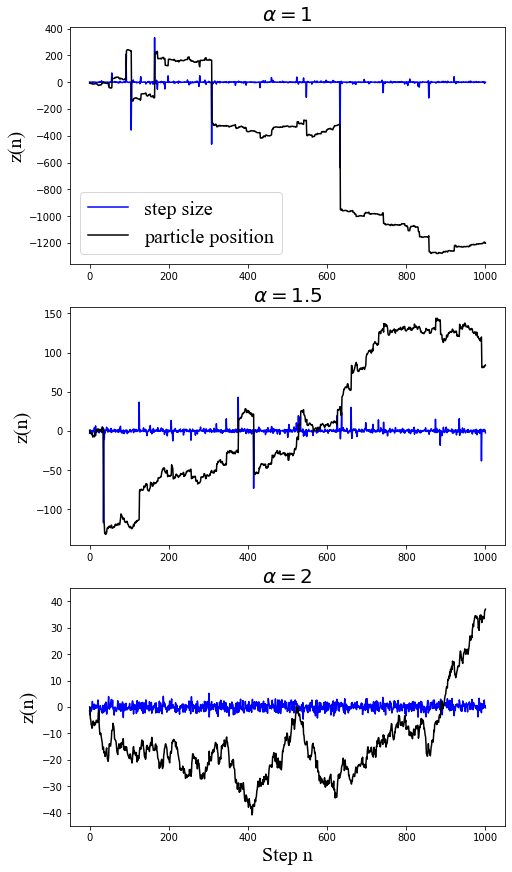}
\caption{\label{fig:path}Monte-Carlo simulation of the trajectories (black) of a particle diffusing at $\alpha=1, 1.5, 2$ and the step-sizes (blue). For $\alpha<2$, the particle undergoes L\'evy flights in each time step, where the pathlength distribution has infinite variance, and the displacement of the particle is dominated by rare large steps. At $\alpha=2$, the pathlengths follow a Gaussian distribution, and the steps contribute more evenly to the overall trajectory.}
\end{figure}

We test our Monte-Carlo algorithm by simulating superdiffusion from delta-function initial conditions for different stability parameters $\alpha$. For simplicity, we choose $\kappa_{\alpha}=10, \Delta t=0.1$, so the distribution of jump lengths at each time step $\Delta t$ all have the same scale, $\gamma = (\kappa_{\alpha}\Delta t)^{1/\alpha} = 1$. We verify directly that we recover the stable distributions seen in Fig. \ref{fig:pdfs}, which are the Green's functions for superdiffusion. We also verify that the net particle displacements scale as $\langle x^2 \rangle \propto t^{2/\alpha}$, as in equation \ref{eq:xsq}, for $\beta=1$. Compared to standard diffusion ($\alpha=2$), the superdiffusive profiles have broad power law tails and thinner, more peaked central regions, i.e. the distribution of particle displacements has less intermediate values, and gets much broader. The extended tails dominate transport. This is shown in Fig. \ref{fig:path}, with three different realizations of 1D diffusion characterized by $\alpha=1, 1.5, 2$. As $\alpha$ decreases, the probability of a particle taking a rare large step increases. As is clearly apparent, standard diffusion ($\alpha=2$) is dominated by the cumulative sum of many small steps, but for smaller $\alpha$, particle transport becomes increasingly dominated by rare large steps -- hence, the `step-function' like appearance of particle position for $\alpha=1$. This highlights the very different nature of fractional diffusion from the standard case.

\subsubsection{Free Diffusion in Spherical Geometry}
\label{subsec:sphere}

Instead of an initial value problem, we now consider continuous injection of CRs in 3D spherical geometry. In contrast to the previous setup, which results in time-dependent, self-similar solutions, continuous injection allows for steady-state solutions. Spherical diffusion mimics a scenario where steady star formation in a galaxy continuously injects CRs at the halo center, which subsequently diffuses into the CGM. In steady state, CR production is balanced by diffusion away from the source\footnote{As seen from the derivation of equation \ref{eqn:Pcr_alpha}, such a steady state with monotonically decreasing CR abundance at large $r$ is only possible when $n>\alpha$, where $n$ is the dimensionality of the problem, due to the more rapid increase in volume with $r$ for higher dimensions.}, so that: 
\begin{equation}
\dot{E}_{\rm CR} = \int dV \nabla \cdot {\mathbf F}_{\rm CR} = \int 4 \pi r^2 dr \kappa_{\alpha} \frac{\partial^{\alpha}}{\partial x^{\alpha}} E_{\rm CR}.  
\end{equation}
As CRs are ultra-relativistic particles, $P_{\rm CR}\approx E_{\rm CR}/3$. Using $\partial^{\alpha}_{x} E_{\rm CR} \sim E_{\rm CR}/r^{\alpha}$, we obtain: 
\begin{equation}
P_{\text{cr}} (r) =\frac{\dot{E}_{\text{cr}}}{12\pi \kappa_{\alpha}r^{3-\alpha}} \Rightarrow E_{\rm CR} (<r) \propto r^{\alpha}
\label{eqn:Pcr_alpha} 
\end{equation}
which is a straightforward generalization of the case for standard diffusion \citep{butsky23}, where $\alpha=2$, and $P_{\rm CR} \propto r^{-1}$. Here, $E_{\rm CR}(< r) \propto r^{\alpha}$ is the total CR energy contained within radius r. Note the scale-free nature of this steady-state solution. Since $P_{\rm CR} \propto r^{-(3-\alpha)}$, superdiffusion (with $\alpha < 2$) steepens the power-law slope of the CR pressure profile. 

\begin{figure}
\centering
\includegraphics[width=0.95\linewidth]{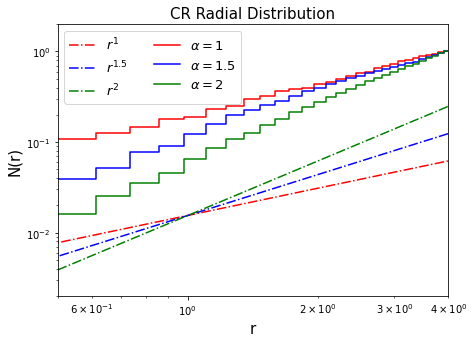}
\caption{\label{fig:3d}The CR radial distributions $N(r)$ from simulations with different stability parameter $\alpha$. The CRs are injected at a fixed rate as a delta point source at the origin. They undergo 3D diffusion in $x,y,$and $z$, and are binned into $r$. The simulations are run until the number of CRs within $r=4$ plateaus and the profiles reach steady states. The scaling of the total number of cosmic rays is consistent with $N(r)\propto r^{\alpha}$, as predicted in Equation \ref{eqn:Pcr_alpha}.}
\end{figure}

We test this expectation in our Monte-Carlo simulations. Particles are injected at a delta point source at the origin and diffuse in 3D (x,y,z).
Given the expected scale-free nature of solutions, we do not adopt physical values of $\kappa_{\alpha}$, but again adopt $\kappa_{\alpha}=10$, $\Delta t=1$, so the scale factor at each time step are all $\gamma = (\kappa_{\alpha} \Delta t)^{1/\alpha} = 1$ in all simulations. For pure diffusion with no energy loss or gain, the number density of CRs follows the same scaling as the energy density. Thus, the number of particles within $r$ should scale as $N_{\text{tot}}(< r)\propto E_{\rm CR}(< r) \propto r^{\alpha}$. As shown in Fig. \ref{fig:3d}, the CR profiles with different $\alpha$ from the 3D simulations follow this scaling. In principle, this change in the radial scaling of the CR pressure profile in the CGM allows one to measure $\alpha$. However, whether this allows for a robust detection of superdiffusion is unclear. For a single central point source, any change in slope is degenerate with a radial change in CR transport properties (such as a scale-dependent diffusion coefficient). See \S\ref{subsec:spatial_dep} for more discussion. 

While free diffusion may be a good model for CR transport in the CGM or ICM, it is not a good model for CR transport close to the Galaxy. The reason is that we have good evidence, from radioactive decay products, that CRs have a relatively short residence time $t_{\rm tot} \sim 10^7$yr in our Galaxy before escaping, certainly much shorter than the age of our Galaxy $t_{\rm MW} \sim 10^{10}$yr. By contrast, the mean age of CRs in our Monte-Carlo simulations does not stabilize but increases monotonically with the simulation runtime. This is because in free diffusion, there is always a finite probability for `old' CRs to diffuse back to the halo center (where CR ages are measured). In Appendix \ref{sec:mean-age}, we show this explicitly (see top panel of Fig. \ref{fig:ages}), and find the analytic scalings. 

\subsection{Diffusion with Absorbing Boundaries}
\label{subsec:absorb}

Despite uncertainties in the CR data and production cross sections, two significant observational constraints for CR propagation in the galaxy are the grammage they accumulated through hadronic interactions, inferred by the secondary (Boron or Beryllium) to primary (Carbon) flux ratio, and the actual age of CRs, estimated from the relative abundance of radioactive components. The inferred short residence times of CRs were first captured in the phenomenological `leaky box' model, where CRs have a an exponentially declining probability to remain in the Galaxy $P(t) \propto {\rm exp}(-t/\tau)$, with $\tau$ as a characteristic residence time. For instance, this could arise if CRs scatter between semi-translucent barriers with some finite probability of escape. The required time for CRs to interact with the ISM in the disk from spallation measurements is $\tau_{d} \sim 3 \times 10^{6}$ years, and the mean residence time CRs spend in the galaxy from radioactivity measurements is $\tau_{\text{tot}} \sim 1.7 \times 10^{7}$ years \citep{munoz77,evoli23}. To reconcile the two timescales, a prevailing hypothesis suggests that CRs cross the central disk several times, where they spallate and acquire their grammage, but spend most of their residence time in the low-density halo, until they eventually diffuse across a chosen halo height $L$ (which can be modeled as an absorbing barrier) and leave the galaxy. This fictitious `halo size' can be envisioned as a boundary where CRs escape due to a change in the magnetic topology, scattering frequency, or transition to advection or streaming. We give an example of the latter in \S\ref{subsec:wind}. The characteristic diffusion time $t \sim L^2/\kappa$ to the absorbing barrier serves as the equivalent of the escape time $\tau$ in the leaky box model, and it can be shown analytically that the distribution of residence times in a diffusion model with absorbing boundaries also obeys an exponential distribution. If only the confinement time $t$ is known, there is a well-known degeneracy between the unknown halo height $L$ and the diffusion coefficient $\kappa$. In principle, if radioactivity and spallation data are jointly analyzed (corresponding to the total confinement time and the time spent in the disk respectively), this degeneracy can be lifted. In practice, the uncertainties are fairly large: for instance, a recent analysis by \cite{maurin22} yielded $L=3.8_{-1.6}^{+2.8}$kpc (using the propagation code USINE, including both data and cross-section uncertainties), and $L= 4.7 \pm 0.6$(data uncertainties)$\pm 2$ (cross-section uncertainties) kpc (using an analytic method). Note also that larger halo sizes $L \sim 5$ kpc are preferred to solve the gradient problem \citep{gabici19}.

Here, we explore the consequences of an absorbing barrier for fractional CR diffusion.

\subsubsection{1D Diffusion with Absorbing Boundaries}
\label{subsubsec:1ddisk}
We first consider 1D simulations, where CRs are assumed to diffuse only away from the disk $z$, and ignore diffusion in the radial direction R. We assume a disk of thickness 0.15 kpc, and add symmetric absorbing boundaries corresponding to halo thickness of $H=5$ kpc. We introduce a point source at the origin injecting particles at a constant rate $\dot{N}$ to mimic the injection of CRs by SNR sources in the galactic disk.  The particles are removed from the system once they reach the halo height $L$. The time particles spent in the disk and their total ages are extracted from a sample of CRs extracted from the disk, since all observations are performed locally within the disk. 

We simulate superdiffusion for different stability parameters $\alpha$. The generalized diffusion coefficient $\kappa_{\alpha}$, which has units of $\text{cm}^{\alpha}\text{s}^{-1}$, is chosen to keep the same mean disk confinement time $t_{\rm disk}$ as we change $\alpha$, so that CRs all acquire the same grammage in different propagation setups. For $\alpha=2$, we choose $\kappa=10^{29}\text{cm}^{2}\text{s}^{-1}$, usually appropriate for GeV protons. The superdiffusive $\kappa_{\alpha}$ for $\alpha=1,1.5$ are rescaled accordingly. The simulations are run for at least 5 $t_{\text{esc}}$, where the escape time $t_{\text{esc}}\sim L^{\alpha}/(\alpha \kappa_{\alpha})$, until CR injection and escape are balanced and the system reaches a steady state. 

\begin{figure}
\centering
\includegraphics[width=0.95\linewidth]{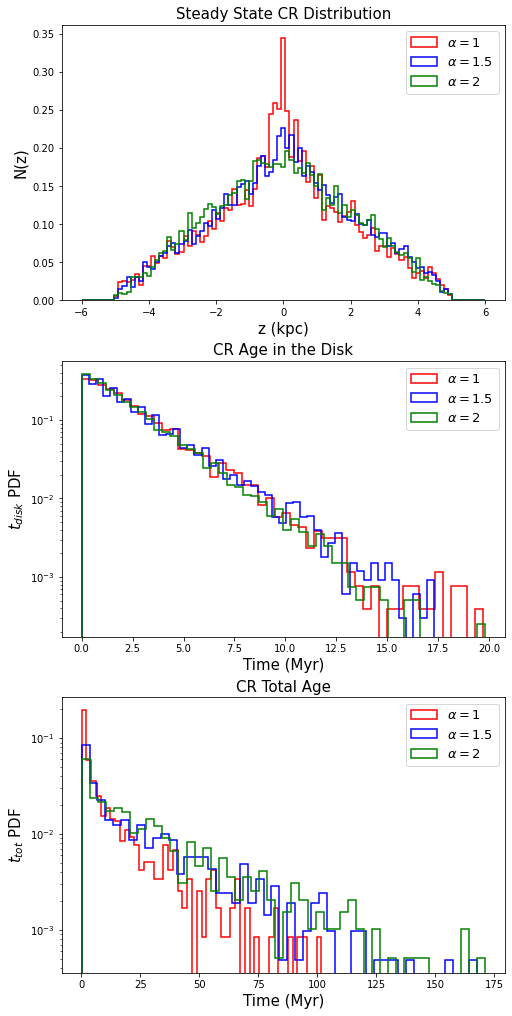}
\caption{\label{fig:1d_abs}(a) The 1D diffusion profile with CRs constantly injected at a delta source at the origin and absorbing boundaries placed at $z=\pm 5$ kpc for different stability parameters $\alpha$. (b) The CR confinement time in the central disk of thickness 0.15 kpc. With absorbing boundaries, the confinement time has an exponential distribution and is independent of the simulation run time. (c) The CR total age distribution. Both of the CR age distributions are extracted from the particles in the central disk, where observational data is available.}
\end{figure}

The results are shown in Fig. \ref{fig:1d_abs}. Perhaps surprisingly, given the noticeably different CR profiles we encountered in the free diffusion case, the CR profiles are very similar and all converge to the same triangular profile (top panel of Fig. \ref{fig:1d_abs}). The only difference is an excess of particles close to the origin for low $\alpha$, which stems from the high peaks at the center of the stable distributions with low $\alpha$ (Fig. \ref{fig:pdfs}). The triangular profile itself can be recovered straightforwardly as an analytic solution of the steady-state standard diffusion equation with the boundary conditions $P_{\rm CR}=0$ at $|z|=L$: 
\begin{equation}
\kappa \frac{\partial^{2} P_{\rm c}}{\partial z^{2}} = \dot{N}_{\rm CR} \delta (0), 
\end{equation}
so that $P_{c} = (\dot{N}_{\rm CR}/\kappa) |L-z|$. As shown in the middle panel of Fig. \ref{fig:1d_abs}, simulations of different $\alpha$ all have the same mean time spent in the disk $t_{\rm disk}$ by construction, and also the same exponential distribution. Despite this, the bottom panel shows that different $\alpha$ have somewhat different total CR lifetimes $t_{\rm tot}$. The total ages do not follow exponential distributions perfectly, but have an excess of short lifetimes, especially for low alpha. In all three simulations, the disk-to-total confinement time ratio are around 0.1 (0.2, 0.1, 0.08 for $\alpha=1,1.5,2$ respectively), consistent with the observational constraints from grammage and radioactivity.

How can we understand these results? The convergence to the same profiles arises from the fact that particles have truncated L\'evy flights (TLFs); the absorbing boundaries effectively remove particles taking large steps. At any position $z$, all particles with step-size larger than $L-|z|$ (in the direction toward the nearest absorbing barrier) are removed. When the variance of the step-size distribution is finite, the Central Limit Theorem applies. As discussed in \S\ref{sec:convergence}, CR transport converges to standard diffusion ($\alpha=2$) after a finite number of scatterings. The reason why the profiles and total particle lifetime PDFs differ when $\alpha=1,1.5$ from the $\alpha=2$ case for small $z$ and small $t_{\rm tot}$ respectively is that it takes a finite amount of time $t_{\rm G}$ for CR transport to gaussianize. Freshly injected particles which are younger than $t_{\rm G}$ have a distribution which is more strongly peaked towards the halo center than in standard diffusion, due the stronger central peak in the corresponding stable distribution (Figure. \ref{fig:pdfs}). This stronger central peak also means that escape probability for young CRs is smaller than for the Gaussian case, which biases the $t_{\rm tot}$ distribution towards younger CRs. Since CR transport and escape probabilities evolve as CRs age and transition from super diffusion to standard diffusion, the $t_{\rm tot}$ PDF is no longer a strict exponential. We discuss the convergence to standard diffusion further in Section \ref{subsec:convergence}. Finally, since CRs are well-mixed, the relative amount of time CR spend in the disk and galaxy as a whole $t_{\rm disk}/t_{\rm tot}$ is simply proportional to the fractional abundance of CRs in the disk, i.e. it can be inferred directly from the top panel of Fig. \ref{fig:1d_abs}. We have confirmed directly in our Monte-Carlo simulations that $t_{\rm disk}/t_{\rm tot} \propto N_{\rm disk}/N_{\rm tot}$ for different $z_{\rm disk}/L$ and for different $\alpha$. Since the fractional disk abundance rises at lower $\alpha$ (due to the stronger central peak), $t_{\rm disk}/t_{\rm tot}$ rises for smaller $\alpha$.  

\subsubsection{Diffusion in Disk Geometry with Absorbing Boundaries}
\label{subsubsec:3ddisk}
We now consider superdiffusion in a more realistic cylindrical geometry, which approximates the geometry of our Galactic disk. This takes into account both the radial variation in CR injection and radial diffusion. It also allows us to address whether superdiffusion can solve the gradient problem, an outstanding problem in Galactic CR transport. The gradient problem arises from the fact that Galactic gamma-ray emission is fairly flat as a function of galactocentric radius R, in contrast with model predictions, which tend to be peaked towards the center due to centrally concentrated star formation and CR production\footnote{A related aspect is that despite these flat gamma-ray profiles, there is a gradient in the CR proton index, which ranges from $E^{-2.3}-E^{-2.5}$ in the innermost 5 pc, to a steeper $E^{-3.0}$ at larger radii \citep{recchia16}.}. Due to the dominance of much larger path lengths in CR transport, one might hope that superdiffusion can decouple the CR profile from the source profile more than standard diffusion, and produce a flat gamma-ray emission profile. 

We run Monte-Carlo simulations where CRs are assumed to be injected by supernovae that roughly follow an exponential radial distribution \citep{stecker77}:
\begin{equation}
Q(\rho)=Q_{0}\rho^{1.2}\text{exp}(-6.44\rho)
\end{equation}
where $\rho = R/L_{R}$. We implement diffusion in Cartesian coordinates (x,y,z) (where diffusion is assumed to be isotropic, $\kappa_x = \kappa_y = \kappa_z$), and bin particle positions via their radial $R$ and vertical distance $z$. The generalized coefficients are chosen to be the same as those used in the 1D tests.
The particles are removed from the system once they reach $L_{z}=5$ kpc or $L_{R}=16$ kpc; since $L_{\rm z} < L_{\rm R}$, most particles are lost through the vertical absorbing boundary.

\begin{figure}
\centering
\includegraphics[width=0.95\linewidth]{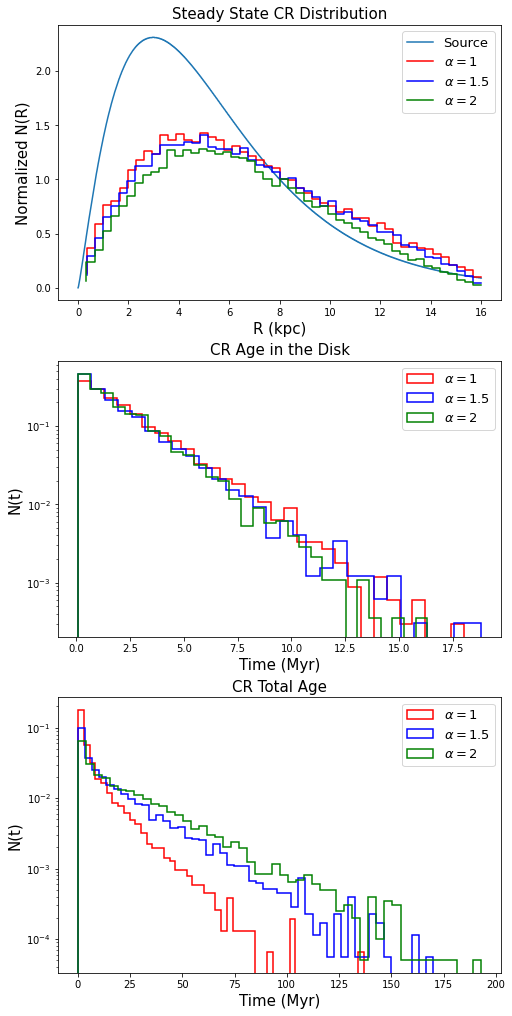}
\caption{\label{fig:2d_abs}
(a) The 3D diffusion (R,z) profiles of CRs injected constantly along R following an exponential distribution of sources $Q(\rho)=Q_{0}\rho^{1.2}\text{exp}(-6.44\rho)$, as in \citet{stecker77} We adopt stability parameter $\alpha=1,1.5,2$, with the mean confinement time in the disk $z= \pm 0.15$ kpc controlled to be similar. Absorbing boundaries are placed both in R and z, with $L_{z}=5$ kpc, and $L_{R}=16$ kpc. The distributions are normalized to unity at $r=8$ kpc. Particles are injected along a fixed radial line (i.e., at a fixed polar angle). Due to cylindrical symmetry, this gives the same result as injecting over the entire disk. (b) The time CRs spent in the central disk for different alphas. The result is similar to 1D diffusion (Fig. \ref{fig:1d_abs}). (c) The CR total age distribution, measured from the particles in the disk.}
\end{figure}

The result are shown in Fig. \ref{fig:2d_abs}, where the CR profile $N(R)$ and particle ages $t_{\rm disk}, t_{\rm tot}$ are measured in the central disk region, where $|z| \leq 0.15$ kpc. The CR distribution is normalized so that $N(R)=1$ for $R=8$ kpc. The profiles in $z$ are all similar to the triangular profiles in 1D. From the top panel, we see that hopes of superdiffusion solving the gradient problem are dashed: the profiles $N(R)$ for $\alpha=1,1.5,2$ are all remarkably similar, with a shape which is somewhat flatter than the source distribution, but inconsistent with observations, which are significantly flatter. With absorbing barriers, the profiles with superdiffusion converge to the same profiles as standard diffusion, for the same reasons as in 1D (\S\ref{subsubsec:1ddisk}): absorbing barriers result in truncated L\'evy flights with finite variance, which therefore gaussianize. As in investigations of the gradient problem with isotropic $\alpha=2$ diffusion, extremely thick halos ($L_{z}/L_{R} \ge 0.75$) and large diffusion coefficients are required to approach the observed gamma-ray distribution, though still not able to match it \citep{webber92}. However, such large $L_{\rm z}$, comparable to the radius of the Galaxy, is not consistent with the short inferred particle escape times.
Proposed solutions to the gradient problem typically involve anisotropic diffusion, which can have galactocentric radial dependence and/or contributions from winds or streaming \citep{reichherzer22}, modifications to the source distribution, or non-linear models of CR transport due to self-generated turbulence \citep{recchia16b}. Since superdiffusion converges to standard diffusion, solutions to the gradient problem will require the same ingredients, and we do not investigate them further in this paper.

Since escape is primarily through the absorbing barrier in $z$, the timescale distributions are similar to that of the 1D case. The time $t_{\rm disk}$ CRs spend in the central disk region $r \le 0.15$ kpc are the same by construction, and they all show an exponential distribution.
The ratio between the time spent in the disk and the total age is also given by $t_{\rm disk}/t_{\rm tot} \sim N_{\rm disk}/N_{\rm tot} \sim (0.3,0.12,0.08)$ for $\alpha=1.0,1.5,2$, where $N_{\rm disk}$ is now integrated over all $R$. Note that while it shows the same trends, $t_{\rm disk}/t_{\rm tot}$ is now a stronger function of $\alpha$ than in the 1D case. This is because CRs can travel at oblique angles which allow for longer flight paths -- i.e., the cut-off length $l_{\rm max}$ is somewhat longer than in the 1D case. This larger cutoff length $l_{\rm max}$ also means a longer gaussianization time $t_{\rm G}$, and thus a bigger difference in $t_{\rm disk}/t_{\rm tot} \sim N_{\rm disk}/N_{\rm tot}$ between standard and fractional diffusion. 

\subsection{Convergence to Standard Diffusion}
\label{subsec:convergence}
CRs diffusing with different stability parameters $\alpha$ converge to the same spatial distribution if there are absorbing boundaries, either in 1D (\S\ref{subsubsec:1ddisk}) or in 3D disk geometry (\S\ref{subsubsec:3ddisk}), but lead to different power-law distributions if we let them freely diffuse in 3D spherical geometry (\S\ref{subsec:sphere}). Here, we examine this result in more detail. With absorbing boundaries, the distributions where the path-lengths are drawn from are indeed truncated L\'evy stable distributions. The particles are prevented from taking extremely long flights because they are removed from the system if flight paths exceed the boundaries. With an effective truncation, the distribution of flight paths has finite variance, so the CLT applies, and superdiffusion converges to normal diffusion after a large number of scatterings. Absorbing boundaries at $L$ differ from imposing a fixed pathlength cutoff $l_{\text{max}}$, since the cutoff is a function of particle position. But they also lead to finite variance and convergence to normal diffusion. As discussed in \S\ref{sec:convergence}, the timescale for gaussianization is of order the time $t \sim l_{\rm max}^{\alpha}/\kappa_{\alpha}$ it takes for particles to diffuse a distance $l_{\rm max}$, meaning that the transport time to the absorbing boundary and the gaussianization time are comparable. Why, then, do particles gaussianize before removal? gaussianization occurs faster than diffusive escape because $l_{\rm max} \lsim L$. The truncation depends on the particle position, and is $l_{\rm max} \sim L$ only for particles at the origin; otherwise it is smaller.

\begin{figure}
\centering
\includegraphics[width=0.95\linewidth]{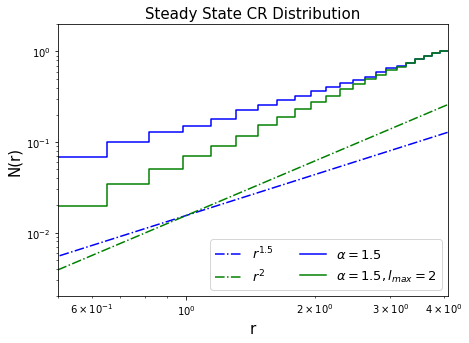}
\caption{\label{fig:3dc} 3D CR superdiffusion profiles with $\alpha=1.5$. The normalized number of particles within r has a scaling $N(r)\propto r^{\alpha} \propto r^{1.5}$ when they take L\'evy flights from stable distributions with infinite variance. When the distributions are truncated to have a finite variance, or when we limit the mean free path to be smaller than $l_{\text{max}}=2$, $N(r)$ converges to the same scaling as normal diffusion, $N(r)\propto r^{2}$.}
\end{figure}

Of course, particle escape is not necessary if pathlengths are limited by some other means (e.g., absorbers or scatters). Consider the original setup of free diffusion in spherical geometry, in a quasi-infinite medium where there is no particle escape (\S\ref{subsec:sphere}), where we found steady-state particle distributions with $N(<r) \propto r^{\alpha}$. If we now truncate L\'evy flights by limiting particle path-lengths below some $l_{\rm max}$, the particle distributions all converge to the profile for standard diffusion, $N(< r) \propto r^2$. We show this in Fig. \ref{fig:3dc}.
Importantly, the removal of particles above a certain age (e.g., CRe which suffer synchrotron or inverse Compton losses) does {\it not} produce `gaussianization', even though naively one might also think of this as a form of `escape'. We have verified this directly in Monte-Carlo simulations. An age limit does not truncate L\'evy flights\footnote{Apart from the light-crossing distance $\sim c t_{\rm age}$ which is so much larger than other scales that convergence to a Gaussian is negligibly slow.}, only the number of steps. Since the pathlength distribution still has infinite variance, the Central Limit Theorem does not apply.

\begin{figure}
\centering
\includegraphics[width=0.95\linewidth]{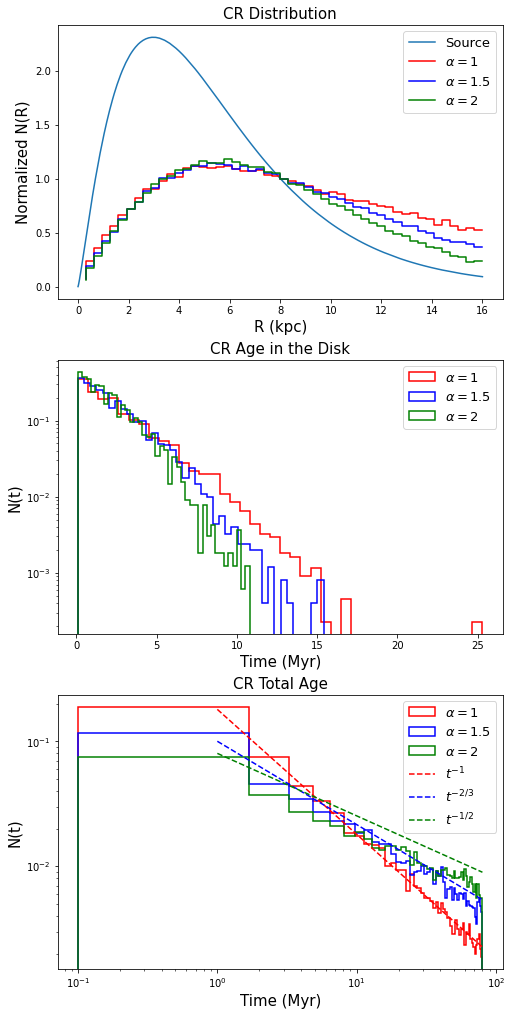}
\caption{\label{fig:free}(a) 3D CR diffusion profiles in disk geometry without absorbing boundaries. The CR source distribution, as well as other parameters, are the same as previous simulations with absorbing boundaries. The simulations are run for $t=40$, where the rms displacement for normal diffusion is expected to be $\sim 5$ kpc. The confinement time in the disk and the total age distribution of particles (which follows $N(t) \propto t^{-1/\alpha}$) are compared in (b) and (c). Both are extracted from particles which reside in the disk, $z=\pm 0.15$ kpc, at the end of the simulation.}
\end{figure}

Removing all absorbing boundaries in the disk setup leads to significantly different spatial distributions. Fig. \ref{fig:free} shows the distributions of CRs diffusing in disk geometry for the same parameters as in \S\ref{subsubsec:3ddisk}, but without absorbing boundaries. Simulations are run for twice the diffusion time to the original boundary location. This choice leads to a similar $t_{\rm disk}$ (i.e. similar grammage) as the original absorbing boundary case. The particle distributions are significantly flatter than the absorbing barrier case. Superdiffusive particles ($\alpha < 2$) exhibit flatter tails at large scales, consistent with the power-law tails of L\'evy distributions. Confinement time in the disk features an exponential cutoff, as the probability of crossing the disk multiple times decreases geometrically. The total age distribution approximately follows $N(t) \propto t^{-1/\alpha}$. This is similar to the $N(t) \propto t^{-3/\alpha}$ age distribution for 3D spherical free diffusion (Fig. \ref{fig:3d}), but in 1D, since disk diffusion is close to 1D diffusion in $z$. Once again, the divergence in mean particle age motivates the necessity of absorbing boundaries.

\begin{figure}
\centering
\includegraphics[width=0.95\linewidth]{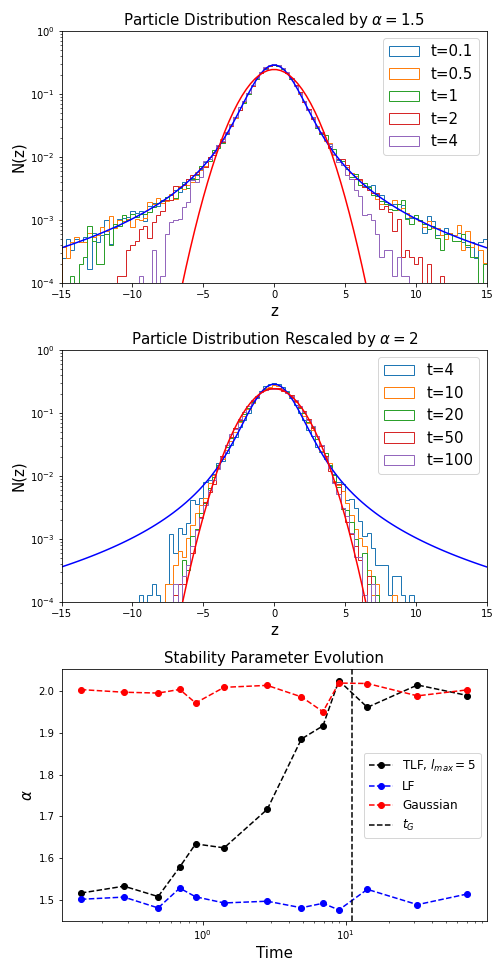}
\caption{\label{fig:rescale} 1D CR distribution at different times for a point source at the origin. The particles undergo truncated L\'evy flights with $l_{\text{max}} = 5$ kpc, where the absorbing boundaries were placed in \S\ref{subsubsec:1ddisk}. The distributions all collapse to the same self-similar shape when they are rescaled by a factor $t_{\text{sim}}^{1/\alpha}$, for $\alpha=1.5$ at early times $t < t_{\rm G}$ in (a) and $\alpha=2$ at late times $t > t_{\rm G}$ in (b). Thus, particles undergo superdiffusion at early times, and converge to standard diffusion at late times. A snapshot $t=4$ of the transition stage is plotted in both scale transformations. In (c), we show the time evolution of the effective $\alpha$, as estimated from the width $\gamma \propto t^{1/\alpha}$. It converges to $\alpha=2$ at the predicted time $t_{G}$.}
\end{figure}

The convergence to standard diffusion can also be shown from the time evolution of particle distributions. To avoid the effect of absorbing boundaries on the distribution shape, we allow particles to diffuse from a delta function point source with $\alpha=1.5$ and the same parameters as in \S\ref{subsubsec:1ddisk} (apart from the absorbing boundary), along with a truncation at $l_{\text{max}} = 5$kpc. In Fig. \ref{fig:rescale}, we observe the collapse of the 1D CR distributions through the scaling transformation $x \rightarrow x/(\kappa_{\alpha} t)^{1/\alpha}$.
Diffusion with stability parameter $\alpha$ is self-similar under this scaling transformation (see equation \ref{eq:G-self-similar}, and note that making histograms of the transformed variable $x/(\kappa_{\alpha} t)^{1/\alpha}$ automatically adjusts the normalization of the PDF, which integrates to unity). Panel (a) represents rescaled distributions with respect to $\alpha = 1.5$, and panel (b) shows that of $\alpha = 2$. It clearly shows that the system transitions from superdiffusion to standard diffusion. In particular, while the distributions initially resemble the $\alpha=1.5$ stable distribution, they progressively deviate in the tails and approach Gaussian shapes due to the CLT. This transition phase is exemplified by the snapshot at $t = 4$. Panel (c) further quantifies this evolution by tracking the effective $\alpha$ obtained from the time-evolution of the distribution width (see below). The transition to standard diffusion $\alpha=2$ agrees well with the predicted convergence time $t_{G}$.

\begin{figure}
\centering
\includegraphics[width=0.95\linewidth]{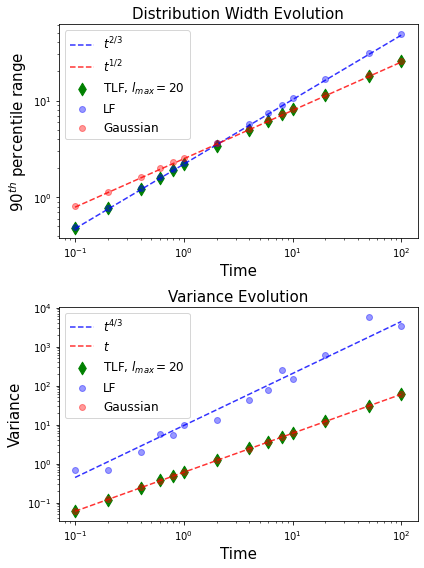}
\caption{\label{fig:width_var}The particle distribution width (defined as the difference between the $95^{\text{th}}$ and the $5^{\text{th}}$ percentile, encompassing $\sim 90\%$ of the particles) (a) and the variance $\langle x^2 \rangle $(b) as a function of time. In (a), the width is expected to scale as $t^{1/\alpha}$. The truncated L\'evy flight (TLF) with $\alpha=1.5$ and $l_{\text{max}}=20$ migrates from the $\alpha=1.5$ to $\alpha=2$ scaling, reflecting a transient period of superdiffusion before gaussianization. In (b), the variance growth is expected to scale as $t^{2/\alpha}$, which is true for standard L\'evy flights. However, the TLF has the same scaling $\langle x^2 \rangle \propto t$ as the Gaussian case at all times, and does {\it not} capture the early period of superdiffusion.}
\end{figure}

We strongly advocate using the power-law evolution in distribution width\footnote{Defined here as the width of the region containing 90\% of the particles, the difference between the $95^{\text{th}}$ and $5^{\text{th}}$ percentile. Note that evaluating the time-dependent width gives the same results as evaluating the time-dependent `probability of return' (Fig. \ref{fig:pr}), which gives the height of the PDF at the origin. Since the product of the width and the height has be constant to preserve normalization, the two have the same scalings, modulo dimensionality (equation \ref{eq:G-self-similar}). However, the width is a more direct and intuitive metric of particle diffusion.} $\gamma \propto t^{1/\alpha}$ to estimate $\alpha$ and diagnose superdiffusion, rather than the commonly practice of using the variance, which in principle scales as $\langle x^2 \rangle \propto t^{2/\alpha}$ (equation \ref{eq:xsq}, for $\beta=1$). The variance is problematic for two reasons: (i) it is a formally infinite for stable distributions. For a finite number of particles, the variance remains finite, but it is dominated by a small number of particles in the tail, and thus is a noisy metric. (ii) For truncated L\'evy flights, the variance always scales as $\langle x^2 \rangle \propto t$, even before the particles have gaussianized and are still super-diffusive. We show this different behavior in the evolution of the width $\gamma \propto t^{1/\alpha}$ and variance $\langle x^2 \rangle \propto t$ for truncated L\'evy flights before gaussianization in Fig. \ref{fig:width_var}. After gaussianization, the width also shows the Gaussian scaling $\gamma \propto t^{1/2}$. We can understand this behavior from the fact that truncated L\'evy flights have variance $\langle x^2 \rangle \propto l_{\rm max}^{2-\alpha} \gamma^{\alpha} \propto t$ \citep{mantegna94,vinogradov10}. Heuristically, the variance is dominated by particles far out in the tail which already `feel' the effect of $l_{\rm max}$, even though most particles in the core are unaffected by the truncation. By contrast, the width $\gamma$ only transitions to Gaussian scalings when most particles are affected by the truncation. {\it This is an important distinction.} Most physical systems (and certainly simulations!) are finite, and subject to truncation. By using the variance as a diagnostic, we suspect that many simulations which report $\langle x^2 \rangle \propto t$ and thus inferred standard diffusion may have missed the opportunity to detect anomalous diffusion during the period $t < t_{\rm G}$. 

\subsection{CR Halo set by Advection and Streaming Dominated Transport}
\label{subsec:wind}

We have previously seen that an absorbing barrier of height $L$ is necessary in standard models to fit CR confinement times with observations. Some means of imposing particle escape once they diffuse a finite distance from the Galaxy, without any chance of diffusing back to the disk, is necessary. We have shown that once such an absorbing boundary is imposed, superdiffusion will gaussianize and converge to standard ($\alpha=2$) diffusion. However, an absorbing boundary is an artificial construct. For a given escape time $t \sim L^{\alpha}/\kappa_{\alpha}$, it is also degenerate with the diffusion coefficient. A potential worry is whether gaussianization depends on the actual physics of how CRs escape. 
Examples of such escape processes include: (i) spatial modulation of the diffusion coefficient, which leads to less scattering and a higher diffusion coefficient at larger radii \citep{ginzburg76}. (ii) A transition to advection or streaming dominated CR flux at large distances, due to an increase in wind speeds or Alfv\'en velocities with distance \citep{bloemen93,breitschwerdt02,evoli18,dogiel20}.

We explore the second mechanism in more detail. Diffusion is a random walk where particles can always return to the origin; hence, they never truly `escape'. By contrast, advection or streaming monotonically carries particles away from the disk. The ratio of streaming/advective flux to diffusive flux is $F_{\rm adv}/F_{\rm diff} \sim (v+v_A) P_c/\kappa \nabla P_c \sim L_c (v+v_A)/\kappa$, which increases away from the disk. The CR scale height $L_{\rm c} = P_c/\nabla P_c$ increases as pressure gradients become more shallow, and the term $(v+v_A)$ also increases: winds accelerate away from the disk, and Alfv\'en speeds $v_{\rm A} \propto \rho^{-1/2}$ increase with decreasing density away from the disk (in 1D geometry where $B$ is roughly constant). Once advection/streaming dominates, the particles have exponentially decreasing probability of diffusing back, so they are effectively ‘absorbed’. We explore simple wind toy models, and show that even with an fuzzy escape boundary which merely facilitates CR escape and does not act as a strict absorbing barrier, a transition to standard diffusion still takes place.  

\begin{figure}
\centering
\includegraphics[width=0.95\linewidth]{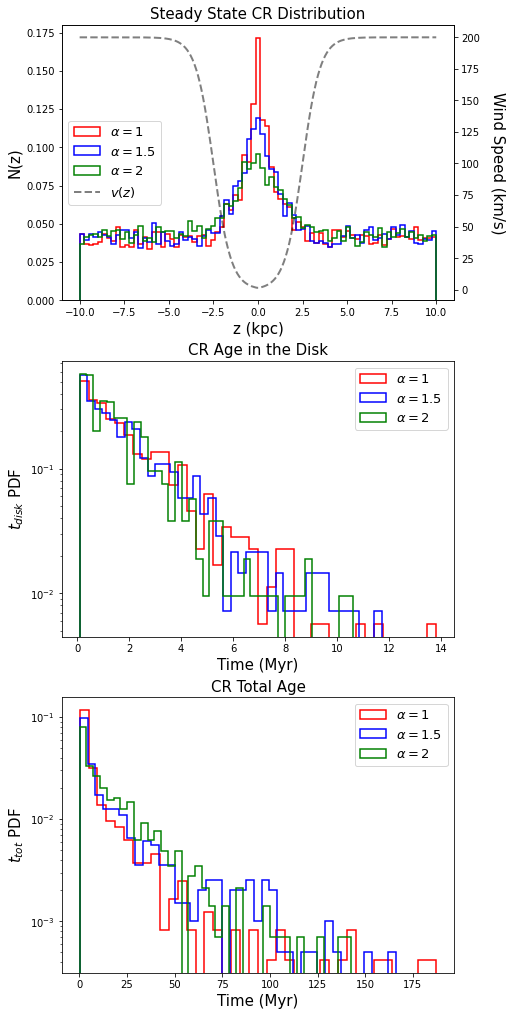}
\caption{\label{fig:tanh_wind}(a) 1D CR distributions with constant injection at the origin and no absorbing boundary, for diffusion with stability parameters $\alpha=1,1.5,2$. Instead, we implement a wind with a tanh function profile starting at $\pm 2.5$ kpc and reaching $\pm 100 \text{km}/\text{s}$ at $\pm 5$ kpc. It quickly dominates over the diffusing flux and effectively removes the particles. The profiles converge, similar to the simulations with absorbing boundaries. (b) Particle confinement time in the disk. (c) Particle total age. Both (b) and (c) are extracted from the central $\pm 0.15$ kpc.}
\end{figure}

First, to create a sharp transition to advection dominated CR transport, we choose a wind speed that rapidly increases at $z=\pm 5$ kpc. In particular, we initialize a tanh wind profile that starts to increase at $z=\pm 2.5$ kpc and reaches $200 \text{km}/\text{s}$ at 5 kpc, a typical speed for a galactic wind. The diffusion coefficients used in all simulations are $\kappa = 10^{29}\text{cm}^{2}\text{s}^{-1}$ for $\alpha=2$, and rescaled for different $\alpha$ to all have the same $t_{\rm disk}$, so they all acquire the same grammage. 
The results are shown in Fig. \ref{fig:tanh_wind}. Modulo the central spike, the particle spatial distributions all collapse to the same triangular profile, for $\alpha=1,1.5,2$ within the effective halo, similar to the absorbing boundary case where they gaussianized (compare with Fig. \ref{fig:1d_abs}). They are uniformly distributed in the advection dominated region, since $N v =$ const and $v=$ const. CR total ages $t_{\rm tot}$ also converge to the same distributions with finite means (with $t_{\rm disk}/t_{\rm tot} \sim 0.1$), as expected if CRs indeed escape (contrast with the bottom panel of Fig. \ref{fig:free}). 

\begin{figure}
\centering
\includegraphics[width=0.95\linewidth]{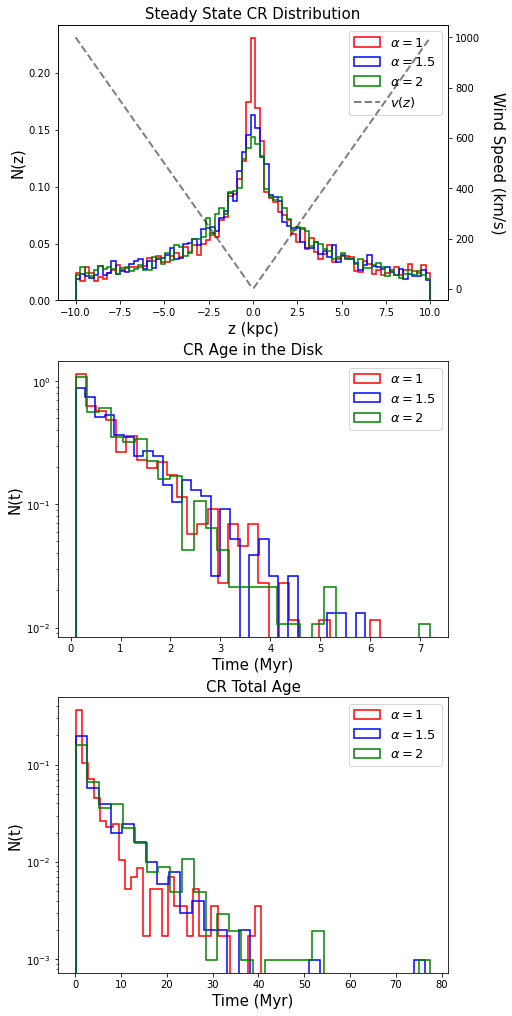}
\caption{\label{fig:lin_wind}(a)1D CR distributions for diffusion with stability parameters $\alpha=1,1.5,2$, with constant injection at the origin and no absorbing boundary. A linear wind begins at the origin and scale up to a few 100 $\text{km}\ \text{s}^{-1}$. It soon dominates the particle flux at $r\sim \pm2.5$ and enables particle escape. Again (modulo the central spike) particles gaussianize and converge to the same profile. Particle confinement time in the disk (b) and particle total age (c) are extracted from a central disk with thickness $\pm 0.15$.}
\end{figure}

\begin{figure}
\centering
\includegraphics[width=0.95\linewidth]{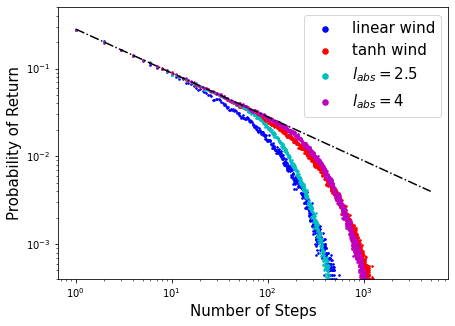}
\caption{\label{fig:wind_pn}The probability of return to the origin for particles that diffuse and advect in a wind, as a function of the number of steps for a tanh profile (red) or linear profile (blue) for the wind velocity. These have probability of return comparable to purely diffusive particles with absorbing boundaries at $l_{\text{abs}} = 2.5, 5$ respectively. The dot-dashed line shows the expectation $P_{\rm n}(0) \propto n^{-1/\alpha}$ for pure diffusion.} 
\end{figure}

We then consider an example where the wind accelerates more gradually: a linear wind profile (Fig. \ref{fig:lin_wind}). We still find that all 3 cases $\alpha=1,1.5,2$ collapse to roughly the spatial distributions. They have similar triangular central profiles as for absorbing boundaries, and flatten when the advective flux is a large fraction of the total flux. We compare this wind model to the case where we place absorbing boundaries at some distance $l_{\rm abs}$, by comparing the probability of return in the two models. For the absorbing boundary case, once the number of steps exceeds a critical number, the probability of return declines exponentially. The wind case shows very similar behavior, and indeed maps on well to the absorbing barrier case if we place the barrier where $F_{\rm adv}/F_{\rm tot} \sim 0.7$, which corresponds to $l_{\rm max}=2.5, 4$ for the linear and tanh wind respectively (Fig. \ref{fig:wind_pn}). This confirms that winds can indeed act like absorbing barriers to limit particle pathlengths and produce gaussianization. The effective barrier height is given by the position where the probability of return deviates from its power-law scaling and declines exponentially. Note that these simple models do not self-consistently allow for the CRs themselves to drive the winds, and ignore the feedback loop between CR transport and wind driving. This more ambitious task awaits future work. 

\subsection{Spatially Dependent Diffusion}
\label{subsec:spatial_dep}

\begin{figure}
\centering
\includegraphics[width=0.95\linewidth]{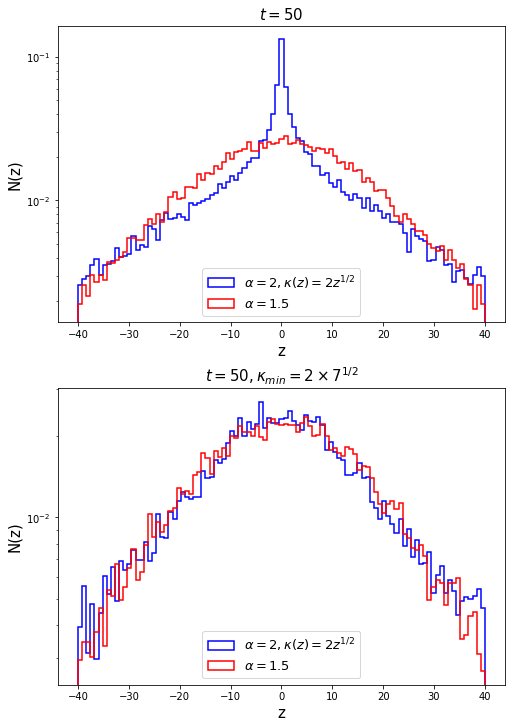}
\caption{\label{fig:spatial_dep}(a) CR profile with spatially varying standard diffusion $\kappa(z) = {\rm max}(\kappa_{\rm min}, 2 z^{1/2})$, compared to superdiffusion with a constant diffusion coefficient, for free diffusion from a delta function at the origin. A minimum value $\kappa_{\rm min}$ is required to allow particles to diffuse away from the origin; the value of $\kappa_{\rm min}$ is increased in (b). For diffusion from the origin, spatially varying standard diffusion closely resembles superdiffusion.}
\end{figure}

\begin{figure}
\centering
\includegraphics[width=0.95\linewidth]{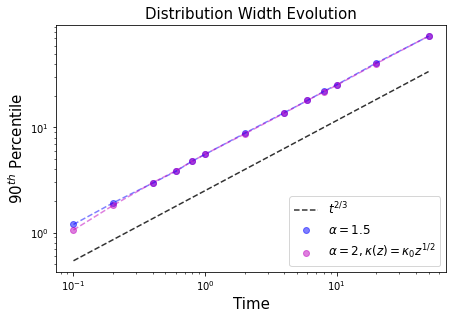}
\caption{\label{fig:spatial_dep_width} Distribution width growth from a delta function at the origin, for spatially varying standard diffusion and superdiffusion with $\alpha=1.5$. By judicious choice of the normalization and scaling of spatially varying standard diffusion, the two have almost identical distribution widths as a function of time.}
\end{figure}

Consider diffusion from a point source. The diffusion time as a function of distance, $t_{\rm diffuse} \sim l^{\alpha}/\kappa_{\rm \alpha}$, can be reproduced by a scale-dependent standard diffusion coefficient with standard diffusion, $t_{\rm diffuse} \sim l^2/\kappa_2(l)$, for $\kappa_2(l) \propto l^{2-\alpha}$ (see also equation \ref{eq:kappa_l}). We show this in Fig. \ref{fig:spatial_dep}. We consider fractional diffusion with stability parameter $\alpha=1.5$, and show that standard diffusion with $\kappa={\rm max}(\kappa_{\rm min}, 2 \sqrt{z})$ gives identical profiles. The floor value $\kappa_{\rm min}$ is just to avoid $\kappa \rightarrow 0$ at $z=0$, which prevents particles from leaving the origin. Similarly, in Fig. \ref{fig:spatial_dep_width}, we show that free fractional diffusion with $\alpha=1.5$ or standard diffusion with $\kappa \propto z^{1/2}$ give distributions whose widths have identical scalings ($\propto t^{1/\alpha} \propto t^{2/3}$) with time. A scale-dependent diffusion coefficient is physically motivated, since we expect the amplitude of magnetic fluctuations which scatter CRs to vary spatially. For instance, if extrinsic turbulence drives these waves, it is expected to vary spatially, since sources of turbulence such as star formation and attendant supernova explosions vary spatially. Similarly, if CRs themselves generate the Alfv\'en waves which scatter them via the streaming instability, the spatially varying CR abundance $n_{\rm CR}(z)$ will change the amplitude of waves. This degeneracy between fractional diffusion and a spatially varying standard diffusion coefficient also complicates efforts to obtain a `smoking gun' signature of fractional diffusion. In \S\ref{subsec:free}, we showed that in steady state, free spherical fractional diffusion from a point source gives power-law profiles $P_{\rm CR} \propto r^{-(3-\alpha)}$. However, a spatially varying standard diffusion coefficient $\kappa \propto r^{2-\alpha}$ would give the same result. For instance, from hydrostatic equilibrium arguments, \citet{butsky23} derives bounds on CR pressure profiles $P_{\rm CR}(r)$ which they interpret to be CR diffusion coefficients $\kappa(r)$ which rise rapidly with radius. \citet{hopkins25} also infer a similar rise in $\kappa(r)$, if excess soft X-ray emission surface brightness profiles are interpreted to be due to inverse Compton emission from CR electrons. These observations could equally well be interpreted as evidence for fractional diffusion. 

Fortunately, this degeneracy between fractional diffusion and spatially varying standard diffusion no longer holds if the CR source distribution and the spatially varying diffusion $\kappa(r)$ no longer have the same symmetry properties. If a point source is offset from the center of symmetry of $\kappa(r)$, then diffusion will be anisotropic (faster along some directions than others), while it will continue to be isotropic for spatially constant fractional diffusion. Thus, when there are multiple sources or some extended single source, the particle distribution will be different for spatially dependent standard diffusion and fractional diffusion. We show this in 1D diffusion for two equal CR point sources at $z=0,5$, at two different times $t=3,50$, where we adopt the same diffusion coefficients as in Fig. \ref{fig:spatial_dep} and \ref{fig:spatial_dep_width}. The profiles differ markedly in both the short and long time limits. In particular, the CR profile for the fractional diffusion case is symmetric about the midpoint between the two sources, $z=2.5$, while this is not true for the standard diffusion case (where $\kappa(z)$ is symmetric about $z=0$). For $t=50$, the separation between the sources is much smaller than the width of the CR distribution, and the two sources can be replaced by a single source of equivalent luminosity at $z=2.5$. We have verified this directly. Thus, homogeneous fractional diffusion and spatially varying standard diffusion can be distinguished if the `center of mass' differs from the center of symmetry of the diffusion coefficient. 

\begin{figure}
\centering
\includegraphics[width=0.95\linewidth]{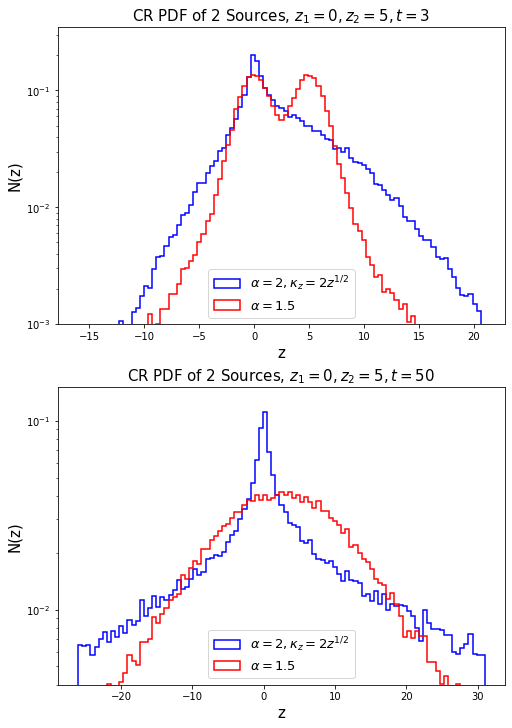}
\caption{\label{fig:2srcs}CR profile for spatially varying standard diffusion $\kappa \propto z^{1/2}$ and constant superdiffusion (with $\alpha=1.5$ when particles are injected at two point sources at $z=0$ and $z=5$. The profiles are different both at early times (a), when the width of each distribution is smaller than the separation, and at late times (b), when the overall distribution evolves like a single source at $z=2.5$. This is because the center of symmetry of the spatially varying diffusion coefficient no longer coincides with the source position.}
\end{figure}

\begin{figure}
\centering
\includegraphics[width=0.95\linewidth]{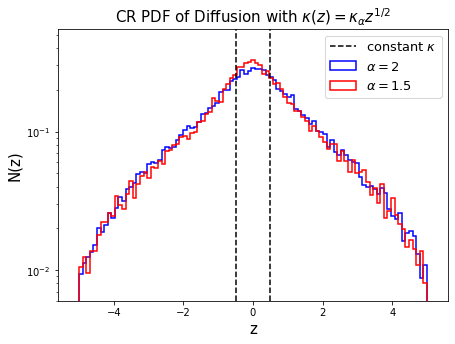}
\caption{\label{fig:spatial_dep_conv}CR distribution for spatially varying superdiffusion coefficient $\kappa_{1.5}(z)=2z^{1/2}$ with an imposed minimal constant value $\kappa_{\text{min}}=2\times0.2^{1/2}$ to diffuse particles out of the origin. Particles are injected at the origin with a constant rate, and are absorbed at $z=5$. Spatially varying superdiffusion converges to a spatially varying standard diffusion with the same z dependence, $\kappa_{\rm eff} \propto z^{1/2}$.}
\end{figure}

Finally, there is no reason why superdiffusion coefficients should be constant. They could also vary spatially,  $\kappa_{\alpha}(r)$, for the same physical reasons that standard diffusion might vary. Previously, in \S\ref{sec:convergence} and \S\ref{subsec:convergence}, we showed that introducing absorbing boundaries or any equivalent cutoff in the pathlength distribution $l_{\rm max}$ causes superdiffusion with constant $\kappa_{\alpha}$ to converge to standard diffusion with a constant equivalent diffusion coefficient $\kappa_{\rm eff} \sim \kappa_{\alpha} l_{\rm max}^{2-\alpha}$. Does spatially varying superdiffusion $\kappa_{\alpha}(r)$ also converge to standard diffusion? We check this this in Fig. \ref{fig:spatial_dep_conv}, where we introduce absorbing boundaries at $z=5$ for a point source at the origin. We find that a spatially varying fractional diffusion coefficient $\kappa_{\rm 1.5} \propto z^{1/2}$ indeed converges to the same profile as a standard diffusion coefficient with the same spatial variation, $\kappa_2 \propto z^{1/2}$. Thus, gaussianization still works for spatially varying fractional diffusion. 

\subsection{Energy Dependence of the Diffusion Coefficient}
\label{subsec:energy}
The measured CR spectrum exhibits a power-law beyond a few GeV. The spectral index is usually attributed to a combination of the CR injection spectrum and diffusive escape losses that further steepens the power-law. In a simplified picture, the steady state energy spectrum can be approximated as $N(E)\sim Q(E)t_{\text{esc}}$, where $Q(E)$ is the source function. Suppose the injection energy spectrum $Q(E)\propto E^{-p}$, and the diffusion coefficient $\kappa_{\alpha}(E)\propto E^{\delta}$, then we obtain $N(E) \propto E^{-p-\delta}$. For example, if we adopt $a=2.7$ from the observed energy spectrum and $p=2$ from Fermi acceleration at strong shocks, then this implies $\delta=0.7$ due to energy dependence of diffusion. On the other hand, the bound on pathlengths could also be energy dependent (e.g., the cross-section of intermittent structures which limit pathlengths could be energy dependent) so that $l(E) \propto E^{\gamma}$. In this case, the escape time for superdiffusion that has converged to standard diffusion is:
\begin{equation}
\label{eqn:tesc_E}
t_{\text{esc}}\sim \frac{L^{2}}{\kappa_{\alpha}(E){l(E)}^{2-\alpha}} \propto E^{-[\delta + \gamma(2-\alpha)]}
\end{equation}
and the overall spectrum $N(E) \propto E^{-a}\propto E^{-p-\delta-\gamma(2-\alpha)}$.  
Given the observed energy spectrum, $N(E) \propto E^{-a}$, all we can determine is the overall energy dependence of the effective diffusion coefficient $\kappa_{\rm eff} (E) = \kappa_{\alpha}(E){l(E)}^{2-\alpha} \propto E^{\delta + \gamma(2-\alpha)}$; even if we knew $\alpha$, we cannot determine $\delta, \gamma$ independently. When $l(E)=L$, i.e. the bound on path-lengths is set by escape from the halo, then this degeneracy is lifted and $t_{\rm esc} \sim L^{\alpha}/\kappa_{\alpha}(E) \propto E^{-\delta}$, the standard formula for fractional diffusion. 

An additional complication is that the halo size itself $L(E)$ could be energy dependent, in which case $t_{\rm esc} \sim L(E)^{\alpha}/\kappa_{\alpha}(E) \propto E^{\alpha\gamma-\delta}$, for $L(E) \propto E^{\gamma}$.  For instance, if $L$ originates from the transition from diffusion to advection or streaming, it inherits this dependence from $\kappa_{\alpha}(E)$. For a simple wind profile $v=v_{0}{(z/z_{0})}^\phi$, and estimating the halo height from the point where $t_{\rm diff} \sim t_{\rm adv}$: 
\begin{equation}
\label{eq:L_k}
    L(E)\sim (\kappa_{\alpha}(E)\frac{{z_{0}}^{\phi}}{v_{0}})^{\frac{1}{{\alpha -1 +\phi}}} \propto E^{\frac{\delta}{(\alpha-1+\phi)}}. 
\end{equation}
Since the exponent of $E$ is always positive, this means that higher energy CRs (which undergo stronger diffusion) have larger escape heights, which is intuitive. If $t_{\rm esc} (E)\propto E^{\omega}$, this gives $\omega = \alpha\delta/(\alpha-1+\phi) - \delta$. Interestingly, for standard diffusion $\alpha=2$ and a linear wind (i.e., constant acceleration such that $v \propto z$, and $\phi=1$), this gives $\omega =0$, independent of $\delta$: particles have the same escape time regardless of their energy, and indeed regardless of the energy dependence of the diffusion coefficient. This seemingly counterintuitive result arises from a cancellation due to the larger halo height $L(E)$ for more energetic particles. We have directly verified it in our linear wind Monte-Carlo simulations. As we vary $\kappa$, $t_{\rm tot}$ remains constant, as predicted, while $t_{\rm disk}$ decreases. Such a model is ruled out by the steepening of the observed spectrum relative to the injected spectrum, which requires the escape time $t_{\rm esc}(E)$ to decrease with energy. But the agreement with Monte-Carlo simulations is a good test of our analytics. 
By contrast, the tanh wind profile, where the wind velocity rises abruptly, has a halo height which is roughly independent of energy. We can measure halo heights in the simulations by measuring the distance at which the probability of return deviates from its power law scaling and starts to decline exponentially (Fig. \ref{fig:wind_pn}). This procedure indeed gives a fixed halo height independent of energy for the tanh wind, and a halo height which scales with energy as in equation \ref{eq:L_k} for the linear wind. 

\subsection{subdiffusion and Continuous Time Random Walks}
\label{subsec:subdiffusion}
\begin{figure}
\centering
\includegraphics[width=0.95\linewidth]{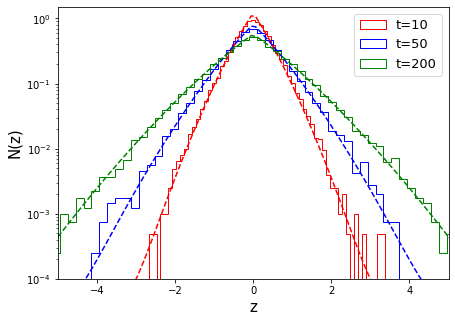}
\caption{\label{fig:foxh}CR spatial distributions from 1D Monte Carlo simulations of subdiffusion with continuous time random walks, where `waiting times' between jumps are drawn from a power-law distribution $P(t) \propto t^{-(1+\beta)}$. Snapshots at $t=10, 50, 200$ (histograms) are compared to the Green’s function (dashed), which is a Fox H function (equation \ref{eq:foxh}). The two show good agreement, indicating our Monte-Carlo simulations of subdiffusion are accurate. This simulation has $\beta=1/2$ and $\alpha=2$, with time step $d\tau=0.1$ and scale factor $\gamma_{z}= 0.2$.}
\end{figure}

\begin{figure}
\centering
\includegraphics[width=0.95\linewidth]{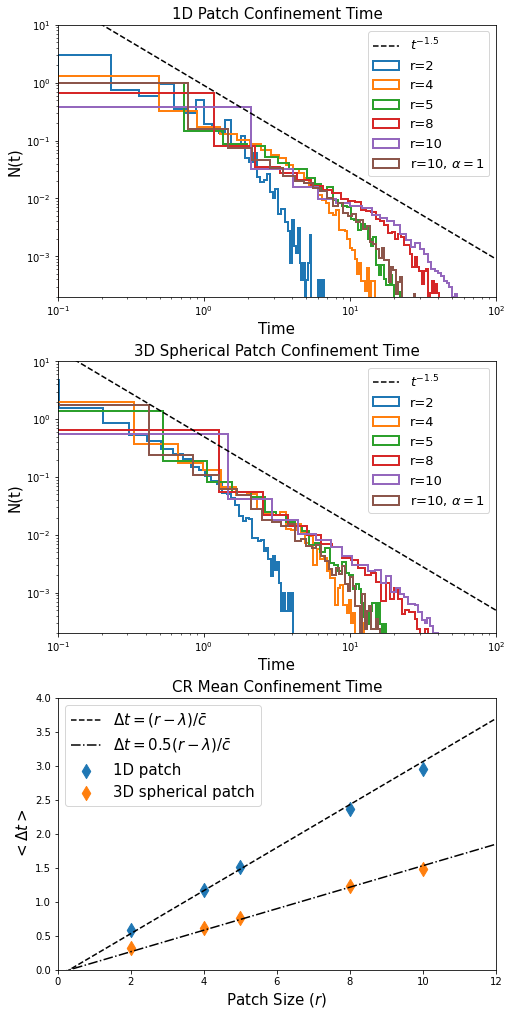}
\caption{\label{fig:sub_time} CR confinement time distributions for CRs incident on patches of size $r$, where they diffuse with diffusion coefficient $\kappa$ in (a) 1D and (b) 3D. Both track the particles entering the patch and undergoing standard spatial diffusion; short confinement times correspond to reflection from the patch. 
The confinement time distributions all follow the power law $N(t) \propto t^{-1.5}$, with a cutoff at the diffusion time $t_{\rm diffuse} \sim L^{2}/\kappa$. The mean confinement time (c) scales with the light crossing time, with different prefactors in 1D and 3D.}
\end{figure}

\begin{figure}
\centering
\includegraphics[width=0.95\linewidth]{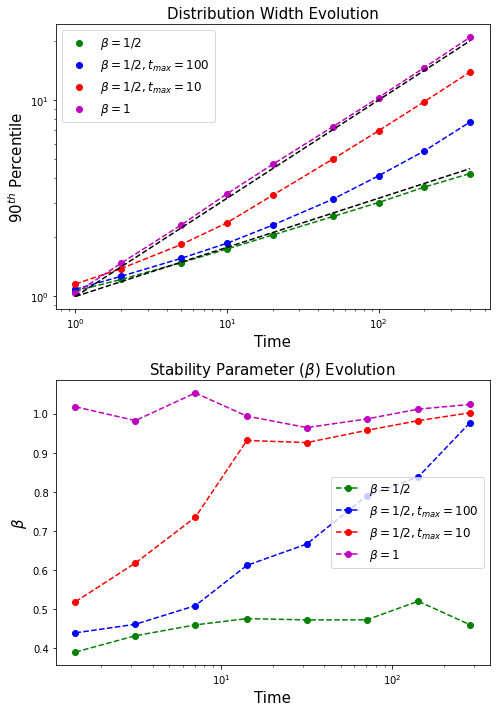}
\caption{\label{fig:sub_conv}(a) CR distribution widths as a function of time for subdiffusion with $\beta=1/2$ and different cutoffs in trapping timescale $t_{\rm max}=10$ (red) and $t_{\rm max}=100$ (blue), compared with the scaling with no cutoff (green) and standard diffusion $\beta=1$(magenta). The scalings for those with truncations change at roughly the maximum trapping time. (b) The corresponding effective $\beta$ parameter evolution, as inferred from the CR distribution width evolution. Truncated subdiffusion with finite trapping time gradually deviates from $\beta=1/2$ and converges to the standard diffusion $\beta = 1$.}
\end{figure}

Besides scattering between the intermittent structures, CRs also spend time diffusing inside the scatterers for various trapping times. We simulate this by a continuous time random walk (CTRW--so called because time is no longer discretized in our simulations, but becomes a continuous random variable) process where particles have power-law waiting times between each step. This leads to a subordinated stochastic process with a time derivative of stability parameter $\beta<1$ in Equation \ref{eq:fractional-diffusion}. The power-law waiting times may stem from the shape and size distributions of the patches, or, as we show here, the stochasticity of the escape process from a patch of given size. In this section, we adopt the assumption of power-law waiting times, draw random times from a power-law distribution, and simulate subdiffusion with steps outlined in Section \ref{sec:monte}. Though we remain agnostic about the detailed characteristics of the intermittent structures that give rise to specific waiting times, we investigate the CR confinement time in patches where transport can be modeled as diffusive, both in 1D and for 3D spheres. For simplicity, we show the results for standard diffusion ($\alpha=2$), and only vary $0< \beta \le 1$, though we have checked that we obtain corresponding results for $\alpha < 2$. 

To test the validity of our Monte-Carlo subdiffusion simulations, we first compare results for free diffusion to the Green’s function for subdiffusion, which is a Fox H function (equation \ref{eq:foxh}). We start with a large number of particles at the origin, and choose $\delta x=0.2$ as the scale factor of the standard Gaussian distribution of step-sizes. This leads to $\kappa_{\beta} = {\Delta x}^{2}/2\tau^{\beta}\Gamma(1-\alpha)$. The result for $\tau=0.1$ and $\beta=1/2$ is compared in Fig. \ref{fig:foxh}. The CTRW results closely resembles the analytic solution.

Next, we find the distribution of confinement times for CRs incident on a patch of size $l_{\rm p}$ with diffusion coefficient $\kappa_{\rm p}$, where scattering is assumed to be stronger in the patch than in the surrounding medium, so that $\kappa_{\rm p} \ll \kappa_{\rm background}$.  We inject particles at the edge of a patch, and record the time spent scattering in the patch. In Fig. \ref{fig:sub_time}, the particle scattering time distributions are shown in panel (a) in 1D and panel (b) for 3D spheres\footnote{In 3D, we choose the incident angle from a uniform distribution in cos$\theta$.} of radius $l$. The distributions have a power-law scaling $N(t) \propto t^{-3/2}$ with exponential cutoffs. The power-law region is more extended for larger patch sizes, as the exponential cutoff corresponds to the diffusion time through the patch, $t_{\rm diff} \sim l_{\rm p} ^2/\kappa_{\rm p}$. In the bottom panel, we show that the mean confinement time scales with the light crossing time $\langle \Delta t_{p} \rangle \sim l_{\rm p}/\tilde{c}$ in 1D, and $\langle \Delta t_{p} \rangle \sim 0.5 l_{\rm p}/\tilde{c}$ for 3D spherical patches, where $\tilde{c}$ is the effective (reduced) speed of light used in the simulations. This agrees  with previous analytic predictions \citep{kotera08,reichherzer23}. 

How can we understand these results? Naively, one might think that the characteristic confinement time should be the diffusion time, $t_{\rm diff} \sim l_{\rm p}^2/\kappa_{\rm p}$. The reason why characteristic times are much shorter is that most particles simply reflect off the clump, without penetrating significantly; only a small fraction actually diffuse through the clump. We can understand the confinement time distribution $N(t)$ from the first-passage time for particle escape and the corresponding survival probability. Consider a 1D setup with a slab of length $l_{\rm p}$, where a particle enters at $z=0$ and must escape through the same boundary by reflection. 
This setup is equivalent to solving the Fokker-Planck equation with an absorbing boundary at $z=0$. In 1D standard diffusion, the solution to this gives a differential distribution of first passage times when the particle is first reflected past the $z=0$ absorbing barrier (e.g., see \citealt{monter24}): 
\begin{equation}
N(t) = \frac{l_{\rm p}}{(2 \pi \kappa_{\rm p} t^{3})^{1/2}} {\rm exp} \left( -\frac{l_{\rm p}^2}{\sqrt{2} \kappa_{\rm p} t} \right). 
\end{equation} 
Thus, for $t \ll t_{\rm diffuse} \sim l_{\rm p}^{2}/\kappa_{\rm p}$, we recover the power-law scaling $N(t) \propto t^{-3/2}$, with an exponential cutoff at $t \sim t_{\rm diffuse}$, when particles diffuse through the patch. In fact, we can go further than that. The Sparre-Andersen scaling implies the survival probability does not depend on the form of the jump distribution as long as it is symmetric, continuous and Markovian \citep{palyulin19}. Thus, the $N(t) \propto t^{-3/2}$ power law scaling also applies for fractional diffusion with stability parameter $\alpha$, as long as $1 < \alpha \le 2$, while for $0< \alpha < 1$, the scaling becomes $N(t) \propto t^{-1 -\alpha/2}$ \citep{palyulin19}. The deviation from the Sparre-Andersen scaling for $\alpha < 1$ arises because the mean jump duration is no longer finite. We have tested these power-law scalings explicitly in our Monte-Carlo simulations. The exponential cutoff now appears at the corresponding diffusion time $t_{\rm diffuse} \sim l_{\rm p}^{\alpha}/\kappa_{\alpha}$. Finally, we can understand the short mean confinement time for the standard diffusion case as follows. To normalize the distribution $N(t) \propto t^{-3/2}$ to unity, we need to know the minimum confinement time $t_{\rm min}$, since $N(t) = t_{\rm min}^{1/2}/(2 t^{3/2})$. The minimum time a CR spends in a patch must correspond to the case where it crosses a mean free path and then is simply reflected out, so that $t_{\rm min} \sim \lambda/c$. On the other hand, as we have seen, the maximum confinement time is of order the diffusion time through the patch, $t_{\rm max} \sim l_{\rm p}^{2}/\kappa$. This gives:
\begin{equation}
\langle \Delta t \rangle \approx \int_{t_{\rm min}}^{t_{\rm max}} t N(t) dt \approx (t_{\rm min} t_{\rm max} )^{1/2} = \left( \frac{\lambda}{c} \frac{l_{\rm p}^2}{\lambda c} \right)^{1/2} \sim \frac{l_{\rm p}}{c}. 
\end{equation} 
where we have used $\kappa_{\rm p} \sim \lambda c$. The mean confinement time in a 3D patch is somewhat shorter, $\langle \Delta t \rangle \sim 0.5 l_{\rm p}/c$, because particles can penetrate the patch at an oblique angle and be scattered closer to the surface. 

The fact that particle confinement times within a single patch already follows a power-law distribution strongly motivates the existence of subdiffusion, and the $N(t) \propto t^{-3/2}$ scaling motivates\footnote{Of course, only for a toy model of a constant diffusion coefficient for a scattering patch.} a choice of $\beta=1/2$; the patch size distribution only modulates the exponential cutoff. The main remaining issue to understand is when and how subdiffusion gaussianizes. In Fig. \ref{fig:sub_conv}, we show the particle distribution width (covering 90\% of the particles) as a function of time for four cases: (i) $\beta = 1/2$, (ii) $\beta = 1/2$ with a maximal waiting time $t_{\text{max}} = 100$, (iii) $\beta = 1/2$ with $t_{\text{max}} = 10$, and (iv) the standard case of $\beta = 1$. We simulate standard diffusion ($\alpha=2$). Indeed, we can see from the top panel that the width scales as $\sim t^{\beta/2}$ when there is no trapping time cutoff, and that it steepens at late times if there is a cutoff. From the bottom panel, we see that the $\beta$ parameter gradually increase for simulations with truncation in trapping time, with $\beta \rightarrow 1$ for $t \gsim t_{\rm G} \sim t_{\rm max}$, i.e. the gaussianization time when the system converges to regular diffusion corresponds to the maximum waiting time. From the high degree of CR isotropy and the inference from grammage and radio-activity measurements that CRs spend most of their time in the volume-filling low density halo, we expect that CR confinement times in our Galaxy to be much larger than trapping times in any particular patch, $t_{\rm tot} \gg t_{\rm max}$. This argues that subdiffusion is unlikely to be important in our Galaxy. 

\section{Discussion}
\label{sec:discussion}

\subsection{When Might CR Fractional Diffusion Matter?}
\label{subsec:whenitmatters}
Our results show that when CR escape from a bounded medium is possible, fractional diffusion will revert to standard ($\alpha=2$) diffusion, on a timescale comparable to the CR escape time. This is because CRs with long path lengths are removed from the system. Truncated L\'evy flights have finite variance, and by the Central Limit Theorem, the distribution of particle displacements approaches a gaussian, and diffusion asymptotes to the standard case obeying Fick's law $F= -\kappa \nabla P_c$. The key is that for many astrophysical systems, particle escape times are short,
generally much shorter than the system lifetime (e.g., compare typical escape times of CRs in our Galaxy $\sim 50 \, {\rm Myr}$ to the Galactic age of $\sim 10 \, {\rm Gyr}$). Hence, diffusion should have gaussianized long ago, and fractional diffusion should be much less prevalent than one might have naively suspected. Note that many small-scale simulations which are used to diagnose anomalous diffusion when CRs scatter in MHD turbulence are not either not run for very long (so CRs may not have a chance to gaussianize), or use periodic boundary conditions (so there is no particle escape). These are important considerations to keep in mind when diagnosing fractional diffusion in the future. 
Alternatively, if there are other ways of limiting CR path-lengths below some maximum $l_{\rm max}$ (e.g., scattering structures of some number number $n$ and cross-section $\sigma$, so that $l_{\rm max} \sim (n \sigma)^{-1}$), gaussianization will also occur. Regardless of the actual mechanism, what this means is that after an initial transient, standard assumptions of Fick's law in CR simulations may indeed be justified. Given the difficulties involved in implementing fractional diffusion into grid codes (\S\ref{sec:monte}), due to the non-local nature of transport, this is good news. 

Consider an application close to the hearts of many extragalactic theorists, CR driven galactic winds. Given that winds are quite different in streaming and diffusion dominated scenarios (e.g., \citealt{wiener17-transport,quataert22-diffusion,quataert22-streaming}), it is natural to ask how fractional diffusion might affect winds. A fully robust answer requires detailed, self-consistent CR hydrodynamic simulations where fractional diffusion is implemented, which would be numerically challenging. However, our results from \S\ref{subsec:wind} suggest that by promoting CR escape via advection (and also via streaming, since the gas density usually falls with distance, increasing the Alfv\'en speed) the wind itself facilitates a transition from fractional to standard diffusion. 

Nonetheless, it would be very interesting to garner observational evidence for fractional diffusion, which would provide important clues about the underlying scattering mechanism. In order for fractional diffusion to be at play, we require $t_{\rm age} < t_{\rm G} \sim t_{\rm esc} \sim R^{\alpha}/\kappa_{\alpha}$, which requires either one to observe the system early on (before gaussianization happens), or to observe a large system where $R \gsim (\kappa t_{\rm age})^{1/\alpha}$ (so that gaussianization never happens during the system lifetime). Potential candidates include: 
\begin{itemize}
\item{{\it CGM, ICM.} The large halos of galaxies and particularly galaxy clusters imply that they can be considered essentially closed boxes where CRs escape only on very long timescales, {\it provided} that CRs continue to scatter strongly and continue to diffuse throughout the volume (e.g., in self-confinement scenarios, the much longer CR densities at large radii mean that CRs scatter much less and approach free streaming; \citealt{wiener13a}). As seen in \S\ref{subsec:sphere}, CR profiles will change if fractional diffusion operates. The challenge is to: (i) find a way to distinguish fractional diffusion from a spatially varying standard diffusion coefficient (as discussed in \S\ref{subsec:spatial_dep}, this is possible if the symmetry properties of CR sources and diffusion differ), (ii) make robust observations of the CR spatial distribution. Hadronic gamma-ray emission is generally too faint to be detectable to large radii, and large-scale synchrotron emission in giant radio halos is degenerate with the B-field profile, as well as the CR electron reacceleration mechanism in merging clusters. A potential candidate might be inverse Compton scattering of CMB photons by high energy ($\gsim$GeV) electrons, which have long ($\gsim$Gyr) lifetimes; this emission comes out in X-ray \citep{hopkins25}. Given the uniform CMB energy density, emission reflects only CR profiles. For instance, we note that the leptonic CR energy density $E_{\rm CR} \propto r^{-2}$ required for inverse Compton emission to explain the observed soft X-ray brightness profiles in the Milky Way and M31 (see Fig 3 in \citealt{hopkins25}) can either be explained by superdiffusion with $\alpha=1$, or a radially dependent standard diffusion coefficient $\kappa_{\rm eff} \propto r$. More detailed modeling would be required to break this degeneracy (\S\ref{subsec:spatial_dep}).} 
\item{{\it Shocks.} It has been claimed that heliospheric shocks show evidence for superdiffusion, from the power-law profiles of electrons accelerated at interplanetary shocks \citep{perri07}, or $\sim $MeV ion profiles at the solar wind termination shock \citep{perri09}, and there has been some work modeling this theoretically \citep{zimbardo13,effenberger24}. While this topic deserves separate study, we note that the downstream flow advects particles away from the shock, allowing for escape, similar to galactic winds. Moreover, since $dn/dE \propto E^{-(1+t_{\rm acc}/t_{\rm esc})}$, observed spectra imply that the acceleration and escape times are comparable, $t_{\rm acc} \sim t_{\rm esc}$. Since particles with large pathlengths will appear far downstream and be removed from the shock, it is not clear if superdiffusion can be maintained, or will gaussianize, like the galactic wind case.}
\item{{\it Pulsar Wind Nebulae.} These are often surrounded by large ($\sim 20$ pc) TeV gamma-ray halos, generated by diffusing electrons and positrons which inverse Compton scatter the CMB. The uniform energy density of the CMB means that gamma-ray profiles unambiguously trace CR profiles, constraining CR transport. Inferred diffusion coefficients are more than two orders of magnitude below Galactic values \citep{abeysekara17}, which is of considerable interest. \citet{wang21} tried using the surface brightness profile to constrain superdiffusion and found that standard ($\alpha=2$) diffusion still provided the best fit, with $\alpha < 1.4$ ruled out at 95\% confidence level. If we take these results at face value, it could reflects the nature of the underlying scattering process, or gaussianization due to particle escape (e.g., via a convective wind, \citealt{hooper17}).}
\item{{\it Cosmic Ray Anisotropy.} While we have studied the effect of gaussianization on CR profiles and concluded that fractional diffusion gives CR profiles identical to standard diffusion once particle escape is taken into account, we have not verified if this is true for CR anisotropy. In principle, the non-local nature of CR transport could make a significant difference here.}

\end{itemize}

\subsection{L\'evy Flights in Radiative Transfer} 
\label{subsec:RT}

As we alluded to in the introduction, CR transport has close parallels with radiative transfer. One might therefore expect that light can undergo L\'evy flights, and that analogs of processes we have discussed should exist in radiative transfer. Here, we briefly touch on a few. 

Optical L\'evy flights have been obtained experimentally by embedding high refractive index scattering particles (titanium dioxide) in a glass matrix, and modulating the local density of these scattering particles by embedding non-scattering glass microspheres with a power-law size distribution \citep{barthelemy08}. By shining a laser on a sample of thickness L, they were able to verify that transmission obeyed the theoretical expectation: 
\begin{equation}
T = \frac{1}{1+A L^{\alpha/2}}
\label{eq:transmission}
\end{equation}
where $A$ is a constant, and $\alpha$ is the stability parameter. Thus, for optically thick materials, $T \propto L^{-\alpha/2}$; they verified this for both $\alpha=1$ (Cauchy) and $\alpha=2$ (Gaussian) distributions. For $\alpha=1$, the average transmitted profile had the sharp cusp and slowly decaying tails characteristic of a Cauchy distribution, compared to the bell curve shape for standard diffusion. Moreover, analogous to the results in this paper, they were able to show in Monte Carlo simulations that after sufficient time, light propagating in the finite-sized sample, which limits pathlengths, transitions from superdiffusion to standard diffusion. 

A situation with elements of both standard and fractional diffusion is resonant line radiative transfer. The astrophysical exemplar is hydrogen Ly$\alpha$ scattering. The Lorentzian scattering cross-section has a Doppler core due to thermal motions and Cauchy tails due to quantum mechanical broadening of the resonance line. The frequency shifts during scattering (absorption and subsequent re-emission of Ly$\alpha$ photons) lead to diffusion in both frequency and space. These are correlated, since scattering cross-sections are smaller and path-lengths are longer as photons diffuse away from line center. The complexity of this process, as well as the fact that we are dealing with Lorentzian rather than strictly L\'evy stable profiles, preclude a straightforward application of the ideas in this paper, such as gaussianization. However, when they become important, the Cauchy tails $\phi(x) \propto x^{-2}$ certainly give rise to `L\'evy-like' behavior. For instance, in very optically thick slabs ($\tau_0 > 10^6$), large jumps dominate: photons do not escape by spatial random walk, but by diffusing far enough into Lorentzian wings so that the medium becomes optically thin and they escape on a `single longest excursion' \citep{adams72}. Such `L\'evy-like' behavior can produce unexpected results. For example, consider Ly$\alpha$ photons incident on a highly optically thick slab with an empty channel -- such as one that is ionized, and therefore does not scatter Ly$\alpha$ photons. Intuitively, we would expect most photons to random walk until they find the channel and escape. Indeed, for standard diffusion, this is true. However, in Monte-Carlo simulations of Ly$\alpha$ radiative transfer, \citet{monter24} find that only a fraction $\sim f_{\rm A}$, where $f_A 
\ll 1$ is the area covering fraction of the channel, escape through the channel; the majority diffuse in frequency and escape through the optically thick slab. The reason is the extended power law tail of scatterings per reflection $\propto N^{-3/2}$, equivalent to the distribution of trapping times $\propto t^{-3/2}$ we see in \S\ref{subsec:subdiffusion}, which enables large frequency shifts and subsequent escape. They also find that transmission scales as $T \propto \tau_{0}^{-1/2}$ (equivalent to equation \ref{eq:transmission} for $\alpha=1$, as appropriate for a Cauchy distribution), rather than the canonical expectation $T \propto \tau_0^{-1}$ appropriate for $\alpha=2$.  

Finally, we briefly note that L\'evy statistics may also be relevant for radio wave scattering, if very small-scale non-Gaussian density fluctuations due to sheets, shocks or ionized boundaries produce a power-law distribution of scattering angles $\theta$ \citep{boldyrev03,boldyrev05,boldyrev06}. These authors argued that L\'evy statistics could explain the observed scaling of temporal broadening of pulsar observations with dispersion measure $\tau \propto {\rm DM}^{4}$, the rapid rise and slow falloff of temporal broadening, and the `cusps' and `halos' seen in scatter broadened images of point-like sources, better than canonical models. Such issues are potentially relevant to CR transport, modulo the connection between density fluctuations which scatter radio waves and the magnetic fluctuations which scatter CRs \citep{kempski24}. We note that the distribution of scattering angles has a cutoff set by the small scale cutoff of density fluctuations, and that the ensuing L\'evy flights could therefore gaussianize. This, and related issues, will be the subject of future work. 

\subsection{L\'evy Flights in CR Acceleration}
\label{subsec:accel}

In this paper, we have focused on fractional spatial diffusion. However, CRs can also diffuse in momentum. This can take the form of `standard' diffusion, where jumps in momentum are small, and superdiffusion, which is dominated by rare large jumps in momentum. For instance, it has been suggested that super-diffusive shock acceleration at the solar wind termination shock can give acceleration times much shorter than standard acceleration \citep{zimbardo13}. Super-diffusive momentum transport has been seen in test particle simulations of CR acceleration in highly turbulent MHD plasmas with strong reconnection, where CRs are accelerated by electric fields at unstable current sheets \citep{isliker17a,isliker17b}. The particles perform large jumps (L\'evy flights) in momentum, and fitting transport coefficients via the classical Fokker-Planck approach (which assumes a Gaussian distribution of step-sizes) fails to reproduce the power-law tails in particle momentum. By contrast, fitting terms in the fractional transport equation (essentially, a version of equations \ref{eq:fractional-diffusion} and \ref{eq:montroll-weiss} in momentum space) and solving it numerically {\it does} reproduce the energy distribution of particles. They assumed $\beta=1$ and fit $\alpha=0.5$ from the momentum jump distribution, consistent with the $N(E) \propto E^{-1.5}$ distribution seen. 

While \citet{isliker17b} allow for particle escape from their simulation box, the system has clearly not gaussianized. Why not? For acceleration to gaussianize, one requires an effective bound on energy jumps $E_{\rm max}$. For instance, particles with $E> E_{\rm max}$ could undergo sharp energy losses, or escape the system and undergo no further acceleration. Instead, the escape time from the simulation box--which is much longer than the acceleration time -- is independent of energy for high energy particles over $\sim 5$ decades in energy; the median number of scatters upon escape also appears independent of energy (see Fig. 3c, 3d of \citealt{isliker17b}). While it would be better to directly examine the fate of particles with large jumps $\Delta E$, the lack of association between high energy and particle escape is consistent with the lack of gaussianization. 

CRs can also undergo sub-diffusive particle acceleration. 3D kinetic particle-in-cell (PIC) simulations of magnetized turbulence, which work with closed boxes (i.e., no particle escape) and neglect energy losses, find power-law energy spectra \citep{zhdankin17,comisso18,wong20}. This might seem at odds with Fermi acceleration, where power-laws develop due to the competition between acceleration and escape. A conventional Fokker-Planck approach indeed predicts pile-up distributions (quasi-Maxwellian distributions with energy concentrated at the maximum momenta allowed by the age of the system), which are not observed. These results were interpreted in the language of continuous time random walks by \citet{lemoine20}, who argued that acceleration `traps' where particles wait in between acceleration events act as an effective escape term. Examples of such traps are would be spatially inhomogeneous acceleration (regions where acceleration is ineffective), or magnetic mirrors (where particles inside the loss cone do not bounce and are not accelerated). Due to the power-law distribution of waiting times, the particles sub-diffuse in momentum. Consistent with our results in \S\ref{subsec:subdiffusion}, \citet{lemoine20} find that if the simulation time is shorter than the maximum trapping time, subdiffusion and power-laws develop. However, on timescales longer than the maximum trapping time, particles gaussianize, and develop the pile-up distributions predicted by the Fokker-Planck equation. 

\section{Conclusions}
\label{sec:conclusion}

We explore the consequences of anomalous diffusion on CR transport, with a particular focus on Galactic CR propagation. In standard diffusion, time-steps have a finite mean (typically with an exponential distribution), and pathlengths have a Gaussian distribution. In anomalous diffusion, time-steps or path-lengths (or both) are drawn from L\'evy stable distributions, and obey the fractional diffusion equation (equation \ref{eq:fractional-diffusion}), with diffusion coefficient $\kappa_{\alpha,
\beta}$, where $[\kappa_{\alpha,\beta}]=L^{\alpha} T^{-\beta}$. Stable distributions are governed by a stability parameter $0 < \alpha \le 2$, which retain the same functional form upon repeated convolution. Well known examples include Gaussians ($\alpha=2$) and Cauchy ($\alpha=1$) distributions. When $\alpha < 2$, they have power law tails $P_{\alpha}(x) \sim x^{-(1+\alpha)}$, which decay more slowly than the exponential tails of Gaussians. This leads to subdiffusion (if time-steps are dominated by rare long waiting times) or superdiffusion (if particle steps are dominated by rare long `L\'evy flights'), where particles diffuse slower or faster than standard diffusion respectively. Both forms are seen in simulations of CR transport in turbulent magnetic fields. Subdiffusion is associated with magnetic traps, while superdiffusion is associated with rapid magnetic field line divergence in MHD turbulence, or potentially with scattering by intermittent structures. While the non-local transport inherent in anomalous diffusion is challenging to implement in grid codes, Monte Carlo simulations similar to those used in Ly$\alpha$ radiative transfer \citep{hansen06} are straightforward to implement. We use these to explore consequences of anomalous diffusion for CR transport. We consider only isotropic diffusion; the more realistic case of anisotropic diffusion is left for future work. 

We first focus on superdiffusion. Our major conclusions are as follows: 
\begin{itemize}
\item{In free diffusion (i.e., when particle escape is unimportant), superdiffusion can substantially change CR profiles. For instance, continuous injection by a point source in a spherical halo gives a steady state CR profile $P_{\rm c} \propto r^{-(3-\alpha)}$.}
\item{However, free diffusion gives excessively large CR ages in the disk, at odds with observations. If CRs escape at absorbing boundaries $H$, so that pathlengths obey $l \lsim H$, superdiffusion gives results identical to standard diffusion. The bound on the CR pathlengths means the distribution now has finite variance, and converges to a Gaussian by the Central Limit Theorem. This occurs on a timescale comparable to the escape time, $t_{\rm esc} \sim H^{\alpha}/\kappa_{\alpha}$, and results in an effective diffusion coefficient $\kappa_{\rm eff} \sim \kappa_{\alpha} H^{2-\alpha}$. Other means of limiting the pathlength to $l \lsim l_{\rm max}$, such as a population of absorbers, would have similar effects. superdiffusion is therefore a transient phase.}
\item{The standard diagnostic for anomalous diffusion is how the variance evolves with time, $\langle x^2 \rangle \propto t^{2/\alpha}$. However, in finite systems where L\'evy flights are truncated, we find $\langle x^2 \rangle \propto t$, even during the initial super-diffusive phase. Instead, we advocate using the distribution width $\gamma \propto t^{1/\alpha}$, which correctly indicates the transition from superdiffusion to standard diffusion. Using this new diagnostic could alter interpretation of previous simulation results.} 
\item{An absorbing boundary is an artificial construct. We should that equivalent gaussianization occurs if CRs are removed by streaming or advection in a wind. The effect halo height is set by the distance where the streaming or advective flux dominates. Thus, after an initial transient, fractional diffusion is unlikely to affect CR-driven winds, which instigate their own gaussianization. By contrast, particle removal by `aging' (e.g., synchrotron or inverse Compton cooling) does {\it not} produce gaussianization, since there is no direct cap on path lengths.}
\item{Anomalous diffusion is still detectable in systems with long CR escape times, such as the CGM/ICM. In systems with sufficient symmetry, fractional diffusion from a point source can be mimicked by a spatially varying standard diffusion coefficient. However, this is in general no longer possible for extended or off-center point sources. Besides the energy dependence of the diffusion coefficient itself, $\kappa_{\alpha}(E)$, the CR energy spectrum can be modified by energy dependence of the limiting pathlength $l_{\rm max}(E)$ or halo height $H(E)$.}
\item{We also simulate subdiffusion. We first simulate scattering off and confinement by intermittent structures, and find trapping time distributions $N(t) \propto t^{-3/2}$ for $1< \alpha\le 2$, and $N(t) \propto t^{-(1+\alpha/2)}$ for $\alpha < 1$, consistent with analytic expectations for first passage times \citep{palyulin19}. While the mean confinement time is only of order the light crossing time, the long tails (which extend to the diffusion time across the trap) allow for significant trapping. Once the simulation time exceeds the maximum trapping time $t_{\rm trap,max}$, CR transport gaussianizes and reverts to standard diffusion. Since observations suggest CRs are isotropic and volume filling rather than localized to small patches, such that the time CRs spend in our Galaxy $t \gg t_{\rm trap,max}$, subdiffusion should be unimportant.} 
\end{itemize}

Besides CR transport, our results have implications for CR acceleration and radiative transfer, which will be the subject of future work. 

\section*{Acknowledgements}
We thank Phillip Kempski, Phil Hopkins, Patrick Reichherzer and particularly Martin Lemoine for stimulating conversations. We acknowledge support from NASA grant 19-ATP19-0205 and NSF grant AST240752. NL acknowledges support from a Paxton fellowship. This research was supported in part by grant NSF PHY-2309135 to the Kavli Institute for Theoretical Physics (KITP). This work was performed in part at the Aspen Center for Physics, which is supported by National Science Foundation grant PHY-2210452. We also acknowledge ACCESS grant PHY240194 for computing support.



\appendix

\section{Mean Age of Particles in Free Spherical Diffusion} \label{sec:mean-age}

\begin{figure}
\centering
\includegraphics[width=0.93\linewidth]{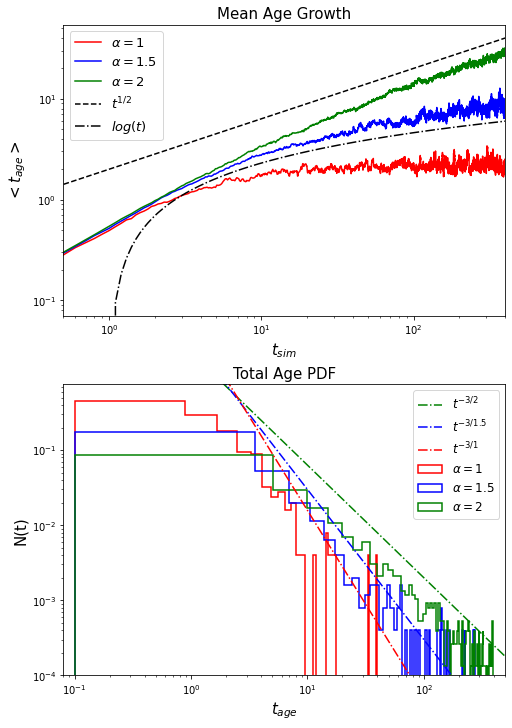}
\caption{\label{fig:ages}(a) The mean CR age from 3D free diffusion simulations increases monotonically with the simulation time $t_{\rm sim}$ either for $\alpha=2$ or $\alpha=1.5$, and plateaus to a constant for $\alpha=1$, consistent with equation \ref{eq:mean-age}. (b) The distribution of CR confinement times within $r=4$, which follow power law distributions $P(t_{\rm age}) \propto t_{\rm age}^{-3/\alpha}$, with a flat region in the center corresponding to newly injected particles.}
\end{figure}

Here, we demonstrate the claim made in \S\ref{subsec:sphere} that for free diffusion, the mean CR age increases monotonically with the age of the system. This is shown in the upper panel of Fig. \ref{fig:ages}, where the mean age diverges in spherical diffusion for $\alpha \gsim 3/2$. Below we show how this behavior can be understood quantitatively. This divergence of CR ages is a strong motivation for introducing absorbing boundaries or escape in Galactic models.
The lower panel of Fig. \ref{fig:ages} shows the distribution of CR ages for spherical diffusion in a fixed region $\Delta r = 4$. 
This age distribution has a power-law tail. The power-law scaling arises from the Green’s function. For standard ($\alpha=2$) 3D diffusion, this is:
\begin{equation}
\label{eqn:3dg} 
G(r,t) = \frac{1}{({4\pi \kappa t})^{3/2}} \Theta(t) e^{-r^{2}/{4\kappa t}}
\end{equation}
where $\Theta(t)$ is the Heaviside step function. Thus, since the final profile is the superposition of Green's functions of CRs released at different times $t$, we have\footnote{Assuming $r \ll (\kappa t)^{1/2}$, which is always true for us, since we extract particle ages in a small window about the origin. } $N(t_{\text{age}}) \propto t_{\rm age}^{-3/2}$ for $\alpha=2$. Though the Green's functions for fractional diffusion do not have analytic forms, they share a similar normalization property that $G(r,t) \propto t^{-3/\alpha}$ for $r^2 \ll (\kappa_{\alpha} t)^{2/\alpha}$ (see equation \ref{eq:G-self-similar}), so we have $N(t_{\text{age}}) \propto t_{\rm age}^{-3/\alpha}$.  

The age distribution has a flat region in the center. This corresponds to newly injected particles. Almost all particles injected within the past $\Delta t \sim (\Delta r)^{\alpha}/\kappa_{\alpha}$ still lie within $\Delta r$. Thus, since the injection rate is uniform, the age distribution is uniform up to $\Delta t$. We have confirmed this by changing the size of the region where we extract particle ages $\Delta r$; the size of the flat region in the age distribution changes accordingly. The age distribution is roughly a convolution of the $N(t_{\text{age}}) \propto t^{-3/\alpha}$ power-law with the uniform distribution of width $\Delta t$. 

The mean particle age then scales as
\begin{equation}
\begin{aligned}
 \langle t \rangle &= \int_{0}^{t_{\text{sim}}} tN(t) \, dt \\
 &\simeq \int_{0}^{t_{\text{flat}}} B t \, dt + \int_{t_{\text{flat}}}^{t_{\text{sim}}} A t^{1-3/\alpha} \, dt \\
 &= \frac{1}{2} B t_{\text{flat}}^2 + \frac{A}{(2 - 3/\alpha)} \left[ t_{\text{sim}}^{2 - 3/\alpha} - t_{\text{flat}}^{2 - 3/\alpha} \right]
\end{aligned}
\end{equation}
where A and B are constants. For $t_{\text{sim}} \gg t_{\text{flat}}$, the integral has limits
\begin{eqnarray}
\langle t \rangle &\propto& t_{\text{sim}}^{2-3/\alpha}; \quad \alpha > 3/2 \label{eq:mean-age} \\ 
&\propto& {\rm log}(t_{\rm sim}) \quad \alpha=3/2 \nonumber \\ 
&\propto& t_{\rm flat}^{2-3/\alpha} \approx {\rm const} \quad \alpha < 3/2\nonumber
\end{eqnarray}
This is consistent with the behavior shown in the top panel of Fig. \ref{fig:ages}, where $\langle t \rangle \propto t^{1/2}$ for $\alpha=2$,  $\langle t \rangle \propto {\rm log}(t)$ for $\alpha=3/2$, and $\langle t \rangle \approx$const for $\alpha=1$. 

One might ask if the young age of Galactic cosmic rays could be explained by the last limit, $\langle t \rangle \approx$const for $\alpha < 3/2$. This is unlikely, as: i) Galactic CR propagation is more 1D than spherical, so the mean age would scale as $\langle t \rangle \propto t^{2-1/\alpha}$, and $\alpha < 1/2$ (extreme superdiffusion) is required to produce an asymptotically constant mean age; ii) young CRs with a uniform age distribution would not produce the exponential grammage distribution that we measure. 

\end{document}